\documentclass[10pt, a4paper]{article}

\usepackage[T1]{fontenc}
\usepackage[utf8]{inputenc}
\usepackage{microtype}
\usepackage{lmodern}
\usepackage[english]{babel}

\usepackage{amscd,amsmath,amsfonts,amssymb,amsthm}
\usepackage{bm}
\usepackage{braket}
\usepackage[mathcal]{euscript}
\usepackage{mathrsfs}
\usepackage{mathtools}
\usepackage{tensor}
\usepackage[output-decimal-marker={.}]{siunitx}
\usepackage[only,llbracket,rrbracket]{stmaryrd}
\usepackage{upgreek}

\usepackage{booktabs}
\usepackage{caption}
\usepackage{bmpsize}
\usepackage{multirow}
\usepackage{subfig}
\usepackage{tabularx}
\usepackage[svgnames,dvipsnames,table]{xcolor}

\usepackage{enumerate}
\DeclarePairedDelimiter{\abs}{\lvert}{\rvert}
\DeclareMathOperator{\uimmi}{\mathrm{i}}

\date{}

\begin{document}

\title{\bf Quasi-periodicity and multi-scale resonators for the reduction of seismic vibrations in fluid-solid systems}

\author{G. Carta$^{a,*}$, A.B. Movchan$^{a}$, L.P. Argani$^{a}$, O.S. Bursi$^{b}$ \\
\small{$^a$ Department of Mathematical Sciences, University of Liverpool, UK} \\
\small{$^b$ Department of Civil, Environmental and Mechanical Engineering,} \\
\small{University of Trento, Italy} \\
\small{$^*$Corresponding author; email address: giorgio\_carta@unica.it} }

\maketitle

\begin{abstract}
\noindent
This paper presents a mathematical model for an industry-inspired problem of vibration isolation applied to elastic fluid-filled containers.
A fundamental problem of suppression of vibrations within a finite-width frequency interval for a multi-scale fluid-solid system has been solved.
We have developed a systematic approach employing full fluid-solid interaction and dispersion analysis, which can be applied to finite and periodic multi-scale systems.
The analytical findings are accompanied by numerical simulations, including frequency response analyses and transient regime computations.
\end{abstract}


\section[Introduction]{Introduction}
\label{Intro}

Mathematical modelling of earthquake mitigation in elastic multi-scale structures is an area of high practical importance.
Unfortunately, real-life structures require large-scale three-dimensional transient simulations, which present a computational challenge and often lead to inconclusive results.
Such problems become even more challenging when fluid-solid interaction is involved and elastic deformations are considered as non-stationary.

\begin{figure}[tp]
\centering
\includegraphics[width=0.9\columnwidth,keepaspectratio]{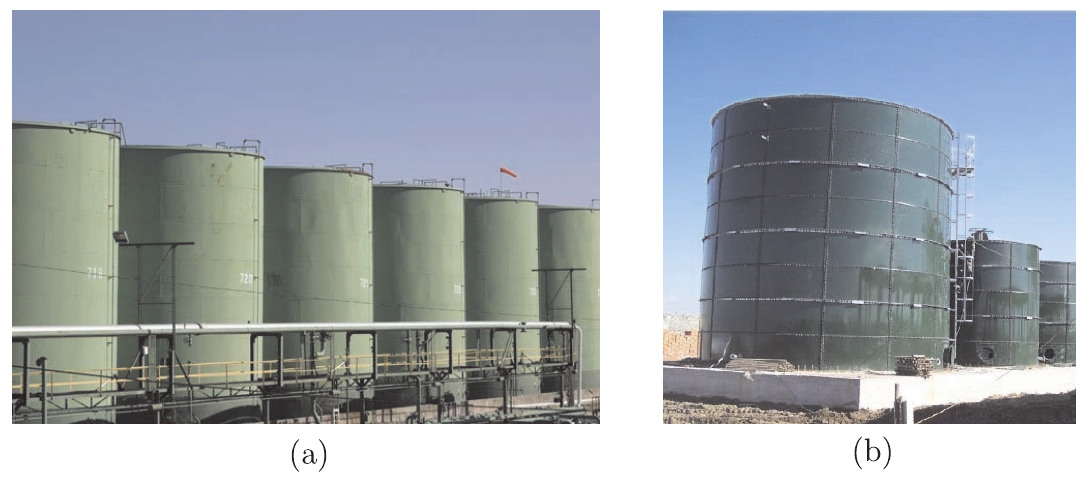}
\caption[Single tank and array of tanks]{(a) Array of tanks in a petrochemical plant (image taken from www.joc.com, accessed on 19/01/2016); (b) cylindrical steel tanks, used to storage water, resting on a concrete foundation (image taken from www.alibaba.com, accessed on 19/01/2016).}
\label{PhotosTanks}
\end{figure}

The purpose of the present paper is to analyse propagation of elastic waves in multi-scale systems of fluid-filled containers and to offer a design leading to suppression of undesirable vibrations.
One class of important applications is in the protection of storage tanks in industrial facilities (see Fig.~\ref{PhotosTanks}) subjected to seismic waves, which can potentially cause serious accidents~\cite{SezWhi2006,Krausmann2010,Bursi2016a}.
Moreover, the proposed design can be applied to reduce the vibrations of different multi-scale structural systems, induced by earthquakes or other dynamic excitations.

In many engineering systems the fluid interacts with a deformable or a moving solid, in which case the coupling between the fluid and the solid needs to be taken into account in the design process.
Fluid-solid interaction problems concerning vibrations of slender structures induced by axial flow are discussed in the comprehensive treatise~\cite{Paidoussis1998,Paidoussis2004}.
In~\cite{BanKun2007} the Distributed Point Source Method was used to derive the ultrasonic field created by ultrasonic transducers in a solid plate immersed in a fluid, and compared the analytical results with the Lamb wave modes visualised experimentally with stroboscopic photoelasticity.
The frequency response of plate structures in contact with a fluid and subjected to an internal harmonic force was investigated in~\cite{Cho2015}.
The resonant frequencies and the associated mode shapes of a rectangular plate lying at the bottom of a container filled with inviscid compressible fluid were calculated in~\cite{LiaoMa2016}.

If the fluid inside a solid has a free surface, sloshing waves are generated when the system is subjected to a dynamic excitation.
Housner~\cite{Housner1957, Housner1963} and other early investigators proposed to model a fluid-filled tank as a mechanical system with masses and springs, where the container flexibility and the sloshing of the liquid free surface are neglected.
The frequencies of sloshing waves can be calculated analytically if the solid container is assumed to be rigid~\cite{Ibrahim2005}.
On the other hand, when the container is elastic, approximate formulations are usually employed (see, for example, \cite{Haroun:1983, Veletsos:1984}).
Alternatively, numerical or experimental investigations can be conducted.
In~\cite{Jiang2014} an experimental sweep test to determine the lower frequencies of sloshing waves in a tank with a rectangular base was performed, considering both thick (rigid) and thin (elastic) walls, and it was found that the resonant frequencies are very close to each other.
In~\cite{PalBha2010} a meshless formulation based on the Petrov-Galerkin method was proposed to study non-linear sloshing waves in a prismatic container under harmonic base excitation, and a good agreement was with the solutions given in~\cite{Washizu1984} and~\cite{Frandsen2004} was obtained.
In engineering applications, the amplitudes of sloshing waves are usually attenuated by using baffles, as shown in~\cite{Belakroum2010} and~\cite{Wang2012}.

Field observations during past earthquakes~\cite{Steinbrugge-Rodrigo:1963, Niwa-Clough:1982, Manos-Clough:1985, Manos:1991, Hamdan:2000} show that storage tanks subjected to seismic loads can be damaged for different mechanisms: large lateral oscillations, buckling of the tank walls (``elephant foot'' and ``diamond shape'' buckling), uplift of the anchorage system, collapse of the tank roof, as well as failure of the piping system.
These different failure mechanisms are described in the state-of-the-art reviews~\cite{Rammerstrofer-Fischer:1990} and~\cite{Ormeno-Larkin-Chouw:2012}.
Many isolation techniques have been developed to prevent damages of tanks, such as linear elastomeric bearings~\cite{Shrimali-Jangid:2002, Shrimali-Jangid:2004}.

In the present paper we focus the attention on the lateral vibrations of the fluid-filled containers. In order to mitigate these vibrations, we propose to introduce a novel system of high-contrast multi-scale resonators, made of many masses linked by light beams and attached to the fluid container.
This system is designed to re-distribute vibrations in a fluid-solid system within a predefined finite frequency range.
The high-contrast multi-scale resonators can be tuned to serve the required frequency interval by varying the masses or the connecting beams.
The proposed design is different from the conventional Tuned Mass Dampers, which are effective only around one or two predefined frequencies.

The possibility to reduce the vibrations in a finite frequency interval is crucial when the spectrum of the system depends on a random parameter, such as the level of fluid inside the container.
Furthermore, earthquake accelerograms are characterised by a wide Fourier amplitude spectrum in the range $[0,30]\si{\hertz}$.
The main advantage of the isolation system devised in this paper is that it is effective in a wide frequency interval, hence it can be applied to a large range of structures (with or without fluid) that can be excited by different frequencies of vibrations within a predefined range.

\begin{figure}[tp]
\centering
\includegraphics[width=1.0\columnwidth,keepaspectratio]{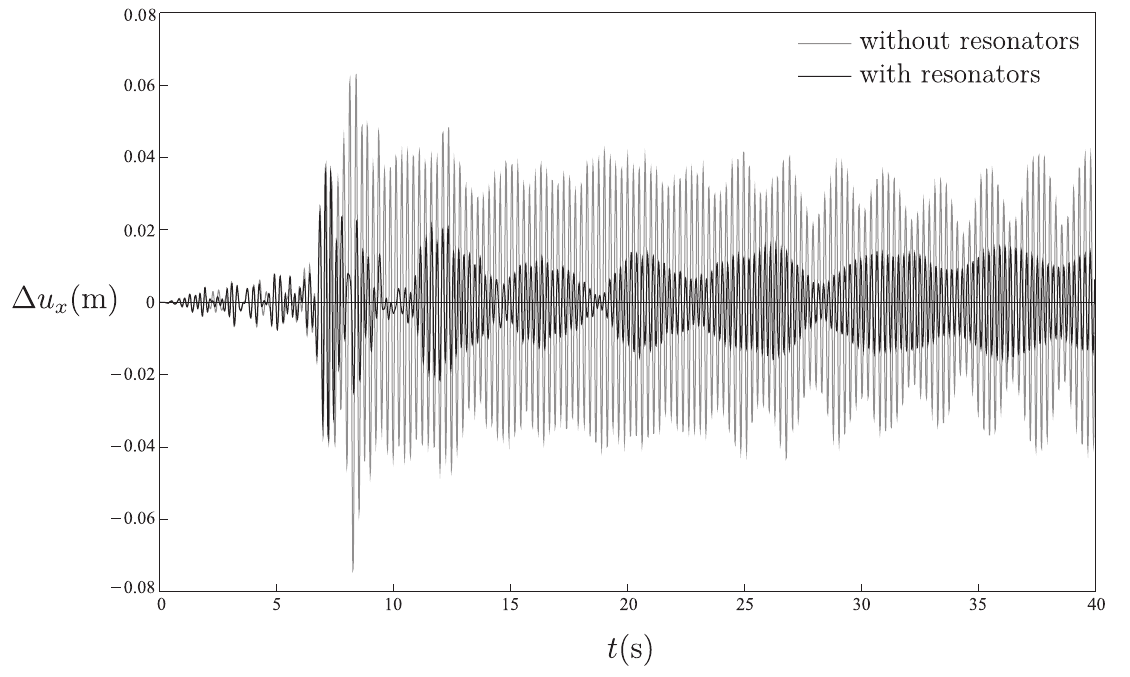}
\caption[Time histories of relative displacements with and without resonators]{Time histories of relative displacements of fluid-filled tanks with resonators (black line) and without resonators (grey line) under the 1994 Northridge NORTHR\_ORR090.AT2 record.}
\label{ExampleEfficiency}
\end{figure}

The example illustrating the efficiency of the proposed design is shown in Fig.~\ref{ExampleEfficiency}, which presents relative displacements of fuel tanks with resonators and without resonators in a real-life earthquake scenario (the seismographic record is taken from the Northridge earthquake of 1994, discussed in Section~\ref{Section3.3}).
It is demonstrated that the modulated vibrations of containers without resonators attain much higher amplitudes than those with the multi-scale resonators proposed in this paper.

We begin by illustrating the use of the multi-scale resonators in the reduction of the vibrations of a three-dimensional cylindrical fuel tank used in a petrochemical plant.
The fluid-solid system and the design of the resonators are described in Section~\ref{Section2}.
In Section~\ref{Section3} we analyse the response of the fuel tank in the frequency domain under a harmonic excitation, as well as the response in the transient regime under real seismic excitations.
We continue by taking a large cluster of connected fluid-filled elastic containers, subjected to externally induced vibrations, as discussed in Section~\ref{Section4}.
A set of many containers is an interesting scenario, considering that in an industrial plant there are areas covered by tank farms (see Fig.~\ref{PhotosTanks}a).
We study the large cluster of containers as a periodic structure and we construct dispersion diagrams, which clearly show existence of stop-bands as well as standing waves in this multi-scale structure.
In Section~\ref{Section5} we assess the effect of the multi-scale resonators on the fluid sloshing waves in the transient domain.
Finally, in~\ref{AppendixA} we discuss several approximations to estimate the resonant frequencies of the combined fluid-solid system, while in~\ref{AppendixB} we present the analytical calculations of the frequencies of sloshing waves.

\section[Multi-scale high-contrast resonators for the seismic protection of fluid-filled tanks]{Multi-scale high-contrast resonators for the seismic protection of fluid-filled tanks}
\label{Section2}

Tanks in an industrial plant are used to store flammable gases or liquids. If the plant is located in a region of high seismicity, they need to resist strong earthquakes without undergoing serious damage.

\begin{figure}[tp]
\centering
\includegraphics[width=1.0\columnwidth,keepaspectratio]{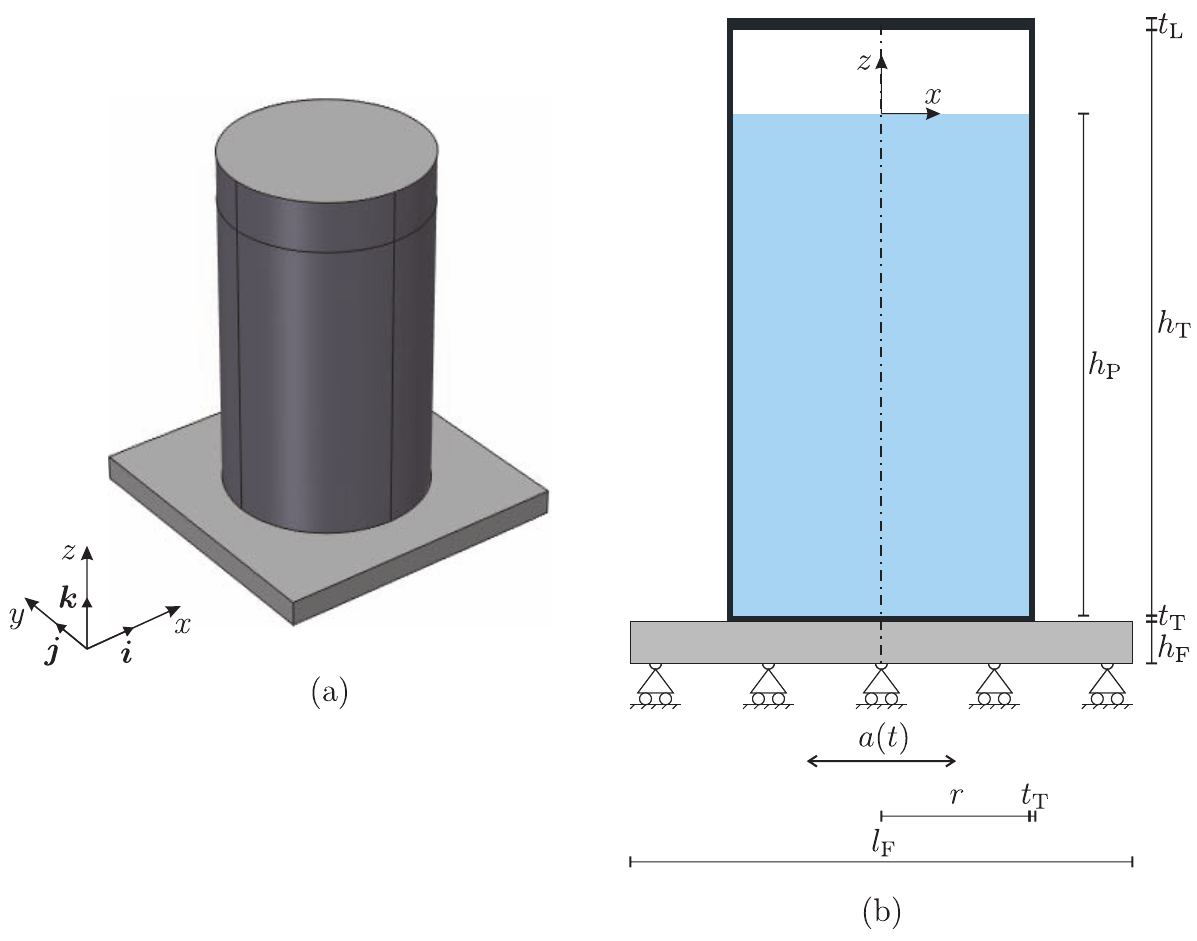}
\caption[Sketch of the three-dimensional fluid filled cylindrical tank]{(a) Three-dimensional cylindrical tank filled with petrol; (b) cross-section of the tank, subjected to a ground acceleration $a(t)$.}
\label{GeometryTank}
\end{figure}

\subsection[Model of a three-dimensional cylindrical tank containing a fluid]{Model of a three-dimensional cylindrical tank containing a fluid}
\label{Section2.1}

A storage tank is generally designed as a steel cylinder with small thickness (see Fig.~\ref{GeometryTank}).
It is covered on top by a circular lid, which is connected to the main structure by pins.
The tank rests on a concrete foundation, which is designed as a square block.
It is filled with petrol, the height of which $h_{\textup{P}}$ is assumed to vary.
Since the bulk modulus of petrol is usually very large, the fluid can be considered as incompressible.

In order to simulate the effects of an earthquake, we assume that the tank is excited by an acceleration $a(t)$ in the $x$ direction, as shown in Fig.~\ref{GeometryTank}.
For simplicity, we assume that the structure is simply supported by the ground in the $z$ direction.

\subsubsection[Governing equations in the transient regime]{Governing equations in the transient regime}
\label{Section2.1.1}

For an incompressible fluid, the continuity equation and the Navier-Stokes equations have the following expressions:
\begin{subequations}\label{EquationsFluid}
\begin{gather}
\nabla \cdot \bm{v}_{\textup{P}} = 0 \, , \label{Continuity} \\
\frac{\partial \bm{v}_{\textup{P}}}{\partial t} + \left( \bm{v}_{\textup{P}} \cdot \nabla \right)\bm{v}_{\textup{P}} + \frac{\nabla p}{\rho_{\textup{P}}} - \frac{\mu_{\textup{P}}}{\rho_{\textup{P}}} \, \nabla^2 \bm{v}_{\textup{P}} = -g \, \bm{k} \, , \label{NavierStokes}
\end{gather}
\end{subequations}
where $\bm{v}_{\textup{P}}$ is the velocity field in the fluid, $p$ is the pressure, $\rho_{\textup{P}}$ and $\mu_{\textup{P}}$ are the density and the dynamic viscosity of the fluid, $g$ is the acceleration of gravity and $\bm{k}$ is the unit vector in the $z$ direction.

The tank and the foundation are studied as elastic continuous systems.
Accordingly, they are governed by the following equations of motion:
\begin{subequations}\label{EquationsSolid}
\begin{align}
\mu_{\textup{T}} \nabla^2 \bm{u}_{\textup{T}} + \left( \lambda_{\textup{T}}+\mu_{\textup{T}} \right) \nabla \nabla \cdot \bm{u}_{\textup{T}} = \rho_{\textup{T}} \frac{\partial^2 \bm{u}_{\textup{T}}}{\partial t^2} & \quad \text{(steel tank)} , \label{EquationsTank} \\
\mu_{\textup{F}} \nabla^2 \bm{u}_{\textup{F}} + \left( \lambda_{\textup{F}}+\mu_{\textup{F}} \right) \nabla \nabla \cdot \bm{u}_{\textup{F}} = \rho_{\textup{F}} \frac{\partial^2 \bm{u}_{\textup{F}}}{\partial t^2} & \quad \text{(concrete foundation)} . \label{EquationsFoundation}
\end{align}
\end{subequations}
$\bm{u}_{\textup{T}}$ and $\bm{u}_{\textup{F}}$ denote the displacement vectors of the tank and the foundation, $\rho_{\textup{T}}$ and $\rho_{\textup{F}}$ are the densities of the tank and the foundation, while
\begin{subequations}
\begin{align}
\mu_{\textup{T,F}}     &= \dfrac{ E_{\textup{T,F}} }{ 2\left(1+\nu_{\textup{T,F}}\right) } \, , \\
\lambda_{\textup{T,F}} &= \dfrac{ E_{\textup{T,F}} \, \nu_{\textup{T,F}} }{ \left(1+\nu_{\textup{T,F}}\right) \left(1-2\,\nu_{\textup{T,F}}\right) } \,
\end{align}
\end{subequations}
are the Lam\'e constants for the tank (subscript \lq T') and the foundation (subscript \lq F'). In the above equations, $\nu_{\textup{T,F}}$ indicate the Poisson's ratios, while $E_{\textup{T,F}}$ are the elastic moduli, which take into account material dissipation as detailed in Section~\ref{Section3.1}.

The boundary conditions of the fluid-solid system are the following:
\begin{subequations}\label{BoundaryConditionsTransient}
\begin{align}
\bm{v}_{\textup{P}} &= \frac{\partial \bm{u}_{\textup{T}}}{\partial t} \, , \quad \bm{\sigma}_{\textup{T}} \, \bm{n} = -p \bm{n} & & \text{at the fluid-solid interfaces} , \label{BoundaryConditionsTransient1} \\
p &= p_0 & & \text{on the fluid free surface} , \label{BoundaryConditionsTransient2} \\
\bm{u}_{\textup{T}} &= \bm{u}_{\textup{F}} & & \text{at the tank-foundation interfaces} , \label{BoundaryConditionsTransient3} \\
u_z^{\textup{F}} &= 0 \, , \, \frac{\partial^2 u_x^{\textup{F}}}{\partial t^2} = a(t) & & \text{at the bottom of the foundation} , \label{BoundaryConditionsTransient4}
\end{align}
\end{subequations}
where $\bm{\sigma}_{\textup{T}}$ is the stress tensor in the tank, $p_0$ is the atmospheric pressure, $\bm{n}$ is the normal unit vector, while $u_z^{\textup{F}}$ and $u_x^{\textup{F}}$ are the components of $\bm{u}_{\textup{F}}$ in the $z$ and $x$ directions.
In the simulations presented in Section~\ref{Section3.3}, the imposed ground acceleration $a(t)$ is taken from real seismographic records.

Finally, we assume that before the ground acceleration $a(t)$ is applied, the system is at rest.
Therefore, the initial conditions are expressed by
\begin{equation}\label{Initial conditions}
\left.\bm{u}_{\textup{T,F}}\right|_{t=0} = \bm{0} \, , \, \left.\frac{\partial \bm{u}_{\textup{T,F}}}{\partial t}\right|_{t=0} = \left.\bm{v}_{\textup{P}}\right|_{t=0} = \bm{0} \, .
\end{equation}

At relatively high frequencies, the fluid can be modelled as an acoustic medium, characterised by the following equation of motion in the transient regime:
\begin{equation}\label{EquationAcousticTransient}
K_{\textup{P}} \nabla^2 p = \rho_{\textup{P}} \, \frac{\partial^2 p}{\partial t^2} \, ,
\end{equation}
where $K_{\textup{P}}$ is the bulk modulus of the fluid.
In this approximate formulation, the contribution of the velocity is neglected and only pressure waves are assumed to propagate in the fluid.
We will show in~\ref{AppendixA} that the eigenfrequencies of the fluid-solid system evaluated with the acoustic approximation are very close to the natural frequencies obtained from the method based on the linearisation of Eqs.~\eqref{EquationsFluid}, which represents the most accurate approach.
The reason is that the frequencies of sloshing waves are considerably small, as shown in~\ref{AppendixB}, therefore the effect of the fluid velocity can be ignored.
Consequently, in the simulations performed in the transient regime and presented in Section~\ref{Section3.3}, the fluid can be alternatively described by Eqs.~\eqref{EquationsFluid} or by Eq.~\eqref{EquationAcousticTransient}.

\subsubsection[Governing equations in the frequency domain]{Governing equations in the frequency domain}
\label{Section2.1.2}

In the frequency regime, the variables of the problem assume time-harmonic dependence.
In particular, the displacements in the tank and in the foundation are given by $\bm{u}_{\textup{T,F}}(\bm{x},t) = \tilde{\bm{u}}_{\textup{T,F}}(\bm{x}) e^{\uimmi \omega t}$, where $\omega$ denotes the radian frequency.

For what concerns the fluid, we consider small harmonic oscillations around the equilibrium configuration, such that $\bm{v}_{\textup{P}} = \bm{V}_{\textup{P}} + \tilde{\bm{v}}_{\textup{P}} \, e^{\uimmi \omega t}$ and $p = P + \tilde{p} \, e^{\uimmi \omega t}$.
At equilibrium the fluid is quiescent ($\bm{V}_{\textup{P}} = \bm{0}$) and $P$ is the hydrostatic pressure ($P = \rho_{\textup{P}} \, g \abs{z}$).
The linearised equations of the fluid in the frequency regime are expressed by
\begin{subequations}\label{EquationsFluidTH}
\begin{gather}
\nabla \cdot \tilde{\bm{v}}_{\textup{P}} = 0 \, , \label{ContinuityTH} \\
\uimmi \omega \, \tilde{\bm{v}}_{\textup{P}} + \frac{\nabla \tilde{p}}{\rho_{\textup{P}}} - \frac{\mu_{\textup{P}}}{\rho_{\textup{P}}} \, \nabla^2 \tilde{\bm{v}}_{\textup{P}} = \bm{0} \, . \label{NavierStokesTH}
\end{gather}
\end{subequations}
For conciseness, the time factor $e^{\uimmi \omega t}$ is suppressed in the following.

\subsection[Analytical design of high-contrast resonators]{Analytical design of high-contrast resonators}
\label{Section2.2}

In order to reduce the vibrations of the tank induced by earthquakes, we propose to insert a passive system of high-contrast multi-scale resonators.
The idea is to attach to the tank an auxiliary structure where waves coming from an external source can be diverted, thus reducing the oscillations of the main structure, which undergoes smaller deformations and is subjected to lower stresses and displacements.
A similar concept has been applied to bridges and tall buildings~\cite{Brun2013}.

\begin{figure}[tp]
\centering
\includegraphics[width=0.8\columnwidth,keepaspectratio]{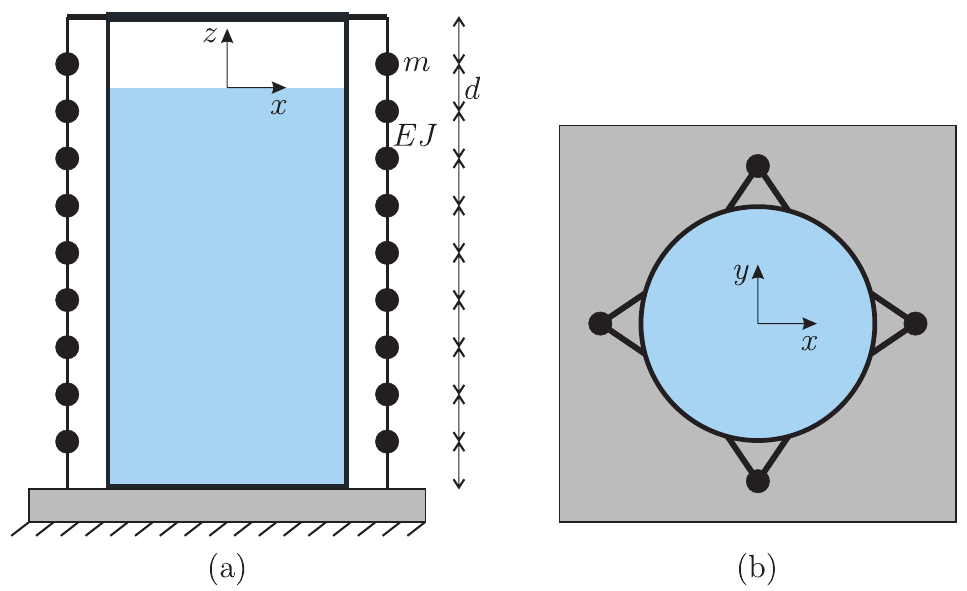}
\caption[Proposed system of resonators]{Proposed system of resonators: front section (a) and top view (b).}
\label{SystemOfResonators3D}
\end{figure}

The system of resonators is made of a large number of masses $m$, distributed at regular distances $d$ in four chains, two of which are located in the $x$ direction and the other two in the $y$ direction (see Fig.~\ref{SystemOfResonators3D}).
The masses are connected by light beams of flexural stiffness $EJ$.
Each chain is clamped at the foundation and at a very rigid truss, that is attached to the top of the tank.

\begin{figure}[tp]
\centering
\includegraphics[width=0.9\columnwidth,keepaspectratio]{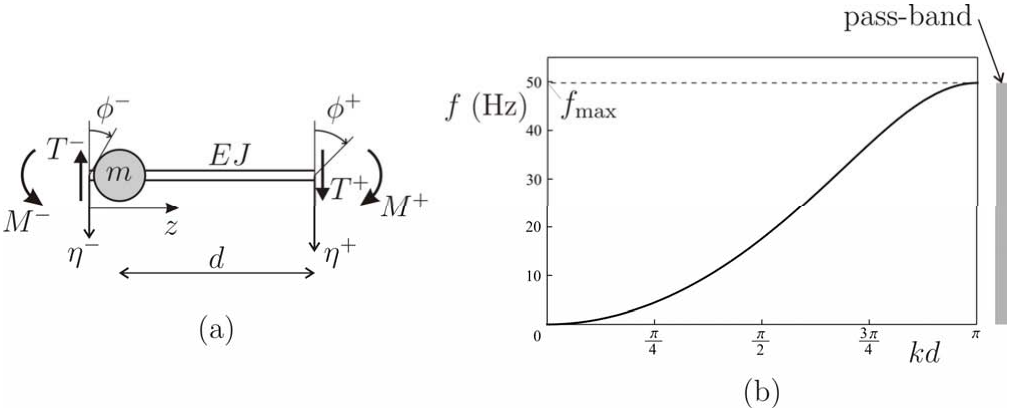}
\caption[Elementary cell of the resonator system and its dispersion curve]{(a) Elementary cell of the system of resonators, analysed as a periodic structure; (b) corresponding dispersion curve (black colour), with the indication of the pass-band (grey colour) (the values of the quantities will be specified in Section \ref{Section3.1.2}).}
\label{DispersionCurve}
\end{figure}

Since the number of masses is large and the beams are very light compared to the masses, the system of resonators is well approximated by a periodic array of masses connected by non-inertial flexural elements.
A repetitive cell of this periodic system is sketched in Fig.~\ref{DispersionCurve}a, where $\eta(z)$ is the transverse displacement, $\phi(z) = \eta'(z)$ is the rotation, $M(z) = EJ \eta''(z)$ is the bending moment and $T(z) = -EJ \eta'''(z)$ is the shear force.
The transfer matrix $\bm{T}$ of this system, which relates the vectors of generalised displacements and forces at the end ($\bm{X}^{+}$) and at the beginning ($\bm{X}^{-}$) of the repetitive cell in the time-harmonic regime, is given by
\begingroup
\renewcommand*{\arraystretch}{1.5}
\begin{equation}\label{TransferMatrix}
\bm{X}^{+} =
  \begin{Bmatrix}
  \eta^{+} \\ \phi^{+} \\ T^{+} \\ M^{+} \\
  \end{Bmatrix}
  = \bm{T} \, \bm{X}^{-} =
  \begin{bmatrix}
  1+\frac{m d^{3} \omega^{2}}{6 EJ} & d & -\frac{d^{3}}{6 EJ} & \frac{d^{2}}{2 EJ}\\
  \frac{m d^{2} \omega^{2}}{2 EJ} & 1 & -\frac{d^{2}}{2 EJ} & \frac{d}{EJ}\\
  -m \omega^{2} & 0 & 1 & 0\\
  m d \omega^{2} & 0 & -d & 1\\
  \end{bmatrix}
  \begin{Bmatrix}
  \eta^{-} \\ \phi^{-} \\ T^{-} \\ M^{-} \\
  \end{Bmatrix} \, .
\end{equation}
\endgroup
The matrix $\bm{T}$ in Eq.~\eqref{TransferMatrix} represents a particular case of the transfer matrix presented in~\cite{Carta2016}, relative to a randomly-perturbed array of translational and rotational masses linked by inertial beams (see Eq.~(5) in the cited paper).

The dispersion relation is derived by imposing quasi-periodicity conditions (or \emph{Floquet-Bloch conditions}) at the ends of the elementary cell, which are expressed as
\begin{equation}\label{FloquetBlochconditions}
\bm{X}^{+} = e^{\uimmi k d} \bm{X}^{-} \, ,
\end{equation}
where $k$ is the wavenumber. The dispersion relation is thus given by
\begin{equation}\label{DispersionRelationGeneral}
\det \left( \bm{T} - e^{\uimmi k d}\bm{I}_{4} \right) = 0 \, ,
\end{equation}
where $\bm{I}_{4}$ is the $4 \times 4$ identity matrix.
Eq.~\eqref{DispersionRelationGeneral} has the following positive solution:
\begin{equation}\label{DispersionRelation}
f = \frac{1}{2 \pi}\sqrt{\frac{6 EJ \left[ 3 + \cos{(2 k d)} - 4\cos{(k d)} \right]}{m d^{3} \left[ 2+\cos{(k d)} \right]}} \, ,
\end{equation}
where $f = \omega/(2 \pi)$ is the frequency measured in~\si{\hertz}.
The dispersion curve obtained from Eq.~\eqref{DispersionRelation} is plotted by a black line in Fig.~\ref{DispersionCurve}b.
The dispersion curve covers the frequency interval $[0,f_{\textup{max}}]$, where $f_{\textup{max}} = 2 \sqrt{3 EJ / (md^{3})}/\pi$.
This frequency interval, indicated in grey colour in Fig.~\ref{DispersionCurve}b, represents the pass-band of the system of resonators, within which waves can propagate; on the other hand, for $f > f_{\textup{max}}$ waves decay exponentially.

All the natural frequencies of the finite system of resonators fall within the pass-band $[0,f_{\textup{max}}]$ for any boundary conditions applied at the ends of the system, as shown in~\cite{Mead1996} and~\cite{Carta2014a,Carta2014b} for other periodic structures.
When the system is excited by a harmonic load having a frequency close to one of the eigenfrequencies of the resonators, the latter start oscillating.
If we introduce a large number of masses $m$, for any frequency within the pass-band $[0,f_{\textup{max}}]$ one eigenmode of the resonators is activated. Therefore, when the system is subjected to an earthquake, all the harmonic components of the seismic load within the range $[0,f_{\textup{max}}]$ are capable of activating the system of resonators, which thus reduce the vibrations of the tank.
We note that the eigenmodes relative to lower eigenfrequencies are more efficient in terms of vibration isolation, because they require more energy to oscillate.

In the next Section, we will show that the system of resonators is indeed capable of mitigating the vibrations of the tank, both in the frequency and in the transient regimes.

\section[Response of fluid-filled tank with resonators to seismic loading - Numerical study]{Response of fluid-filled tank with resonators to seismic loading - Numerical study}
\label{Section3}

\subsection[Geometric and constitutive properties of the fluid-solid system]{Geometric and constitutive properties of the fluid-solid system}
\label{Section3.1}

\subsubsection[Fluid-filled tank]{Fluid-filled tank}
\label{Section3.1.1}

We consider a slender storage tank containing petrol, typically found in a petrochemical plant, which represents one of the case studies of the European project INDUSE-2-SAFETY~\cite{Bursi2016b}.

The tank has radius $r = \SI{4}{\metre}$, thickness $t_{\textup{T}} = \SI{0.006}{\metre}$ and height $h_{\textup{T}} = \SI{14}{\metre}$.
It is made of steel, having Young's modulus ${\bar E}_{\textup{T}} = \SI{190}{\giga\pascal}$, Poisson's ratio $\nu_{\textup{T}} = \num{0.3}$ and density $\rho_{\textup{T}} = \SI{7870}{\kilogram.\metre^{-3}}$.
The covering lid has thickness $t_{\textup{L}} = \SI{0.08}{\metre}$ and has the same constitutive properties as the tank.

The foundation is a square block of side $l_{\textup{F}} = \SI{12}{\metre}$ and height $h_{\textup{F}} = \SI{1}{\metre}$.
It is made of concrete, characterised by a Young's modulus ${\bar E}_{\textup{F}} = \SI{30}{\giga\pascal}$, a Poisson's ratio $\nu_{\textup{F}} = \num{0.2}$ and a density $\rho_{\textup{F}} = \SI{2500}{\kilogram.\metre^{-3}}$.

The properties of petrol are the following: bulk modulus $K_{\textup{P}} = \SI{1.3}{\giga\pascal}$, density $\rho_{\textup{P}} = \SI{750}{\kilogram.\metre^{-3}}$ and dynamic viscosity $\mu_{\textup{P}} = \SI{0.0006}{\pascal.\second}$.

The effect of damping in the system is taken into account by introducing an isotropic loss factor $\zeta = \num{0.03}$.
Therefore, the actual elastic moduli of the tank (subscript \lq T') and the foundation (subscript \lq F') are given by $E_{\textup{T,F}} = {\bar E}_{\textup{T,F}} \left( 1 + \uimmi \zeta \right)$, respectively.

\subsubsection[Multi-scale resonators]{Multi-scale resonators}
\label{Section3.1.2}

We design the resonators to cover the interval $[0,50]\si{\hertz}$.
Accordingly, the parameters of the resonators need to be chosen such that the upper limit of the dispersion curve $f_{\textup{max}} = \SI{50}{\hertz}$, as in Fig.~\ref{DispersionCurve}b.

We divide each chain in~\num{50} equal intervals of length $d = \SI{0.28}{\metre}$ and we place~\num{49} masses at the nodes.
We develop two different designs:
\begin{itemize}
\item \emph{design 1}: we take $m = \SI{304.5}{\kilogram}$ and $EJ = \SI{13119.4}{\newton.\metre^{2}}$, which lead to a total mass of the resonators equal to~\num{10}\% the total mass of the fluid-filled tank (excluding the concrete foundation);
\item \emph{design 2}: the resonators have mass $m = \SI{151.1}{\kilogram}$ and flexural stiffness $EJ = \SI{6943.7}{\newton.\metre^{2}}$, so that the total mass of the resonators is~\num{5}\% the total mass of the fluid-filled tank (without the concrete foundation).
\end{itemize}
We notice that \emph{design 2} represents a more practical and economical solution than \emph{design 1}. The properties of the resonators used in \emph{design 2} can be obtained by using steel disks of thickness $\SI{0.05}{\metre}$ and radius $\SI{0.35}{\metre}$, linked by steel beams of circular cross-section with diameter $\SI{0.029}{\metre}$.

\subsection[Analysis of the fluid-solid system in the frequency domain]{Analysis of the fluid-solid system in the frequency domain}
\label{Section3.2}

First, we determine the dynamic response of the system under external harmonic excitations.
In order to show that the resonators are efficient in a wide frequency interval, we consider different fluid levels in the tank between the extreme situations in which the container is full or empty.

We apply a sinusoidal acceleration in the $x$ direction having amplitude $\tilde{a} = \SI{1}{\metre.\second^{-2}}$ and we compute the horizontal displacements at the top and at the bottom of the tank, indicated by $\tilde{u}_{x}^{\textup{T}}$ and $\tilde{u}_{x}^{\textup{B}}$ respectively.
The numerical simulations are performed with the finite element software \emph{Comsol Multiphysics}\textsuperscript{\textregistered}.
The tank is modelled as a shell to simplify the computations.

\begin{figure}[tp]
\centering
\includegraphics[width=1.0\columnwidth,keepaspectratio]{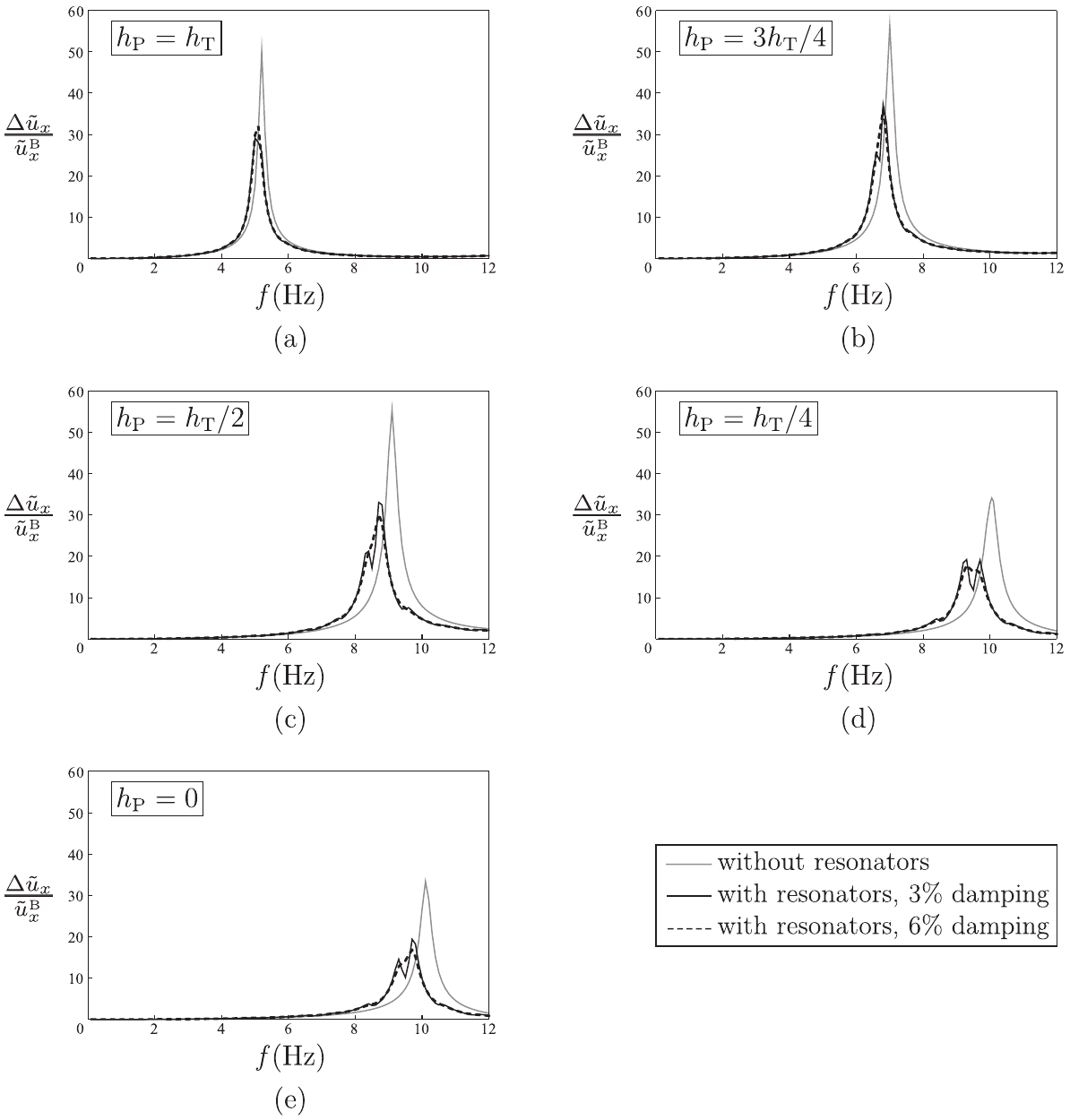}
\caption[Response of the fluid-filled tank in the frequency domain]{Response of the fluid-filled tank in the frequency domain in absence of resonators (grey lines) and when resonators are installed (black lines), for different tank fillings: $100\%$ (a), $75\%$ (b), $50\%$ (c), $25\%$ (d) and empty (e) (\emph{design 1}).}
\label{FrequencyDomain1}
\end{figure}

In Fig.~\ref{FrequencyDomain1} we plot the ratio of the relative displacement $\Delta \tilde{u}_{x} = \tilde{u}_{x}^{\textup{T}} - \tilde{u}_{x}^{\textup{B}}$ to the displacement at the bottom $\tilde{u}_{x}^{\textup{B}}$ for different frequencies, both when the resonators are absent (solid grey lines) and when they are attached to the tank (solid black lines).
We remind that in both cases the isotropic loss factor is~\num{3}\% in the whole structure.
Five different fluid levels have been examined, in particular $h_{\textup{P}} = h_{\textup{T}}$ (inset a), $h_{\textup{P}} = 3 h_{\textup{T}}/4$ (b), $h_{\textup{P}} = h_{\textup{T}}/2$ (c), $h_{\textup{P}} = h_{\textup{T}}/4$ (d) and $h_{\textup{P}} = 0$ (e).
By performing an eigenfrequency analysis, we have checked that the peaks in the response of the system without resonators for the different tank fillings occur in correspondence with the eigenfrequencies of the system associated with antisymmetric modes.

Fig.~\ref{FrequencyDomain1} shows that the resonators reduce and shift the maximum amplitudes of the system response for all the different fluid levels considered.
In most of the cases two peaks are generated, which have considerably smaller amplitudes than those exhibited by the system without resonators.

The grey dashed lines in Fig.~\ref{FrequencyDomain1} represent the responses of the system endowed with resonators when the isotropic loss factor in the resonators is increased to~\num{6}\%.
Obviously, for most of the fluid levels investigated the larger is the damping, the smaller are the vibration amplitudes.
However, the decrease in the peaks of the structural response is not so relevant to justify a damping increment in the system of resonators.

\begin{figure}[tp]
\centering
\includegraphics[width=1.0\columnwidth,keepaspectratio]{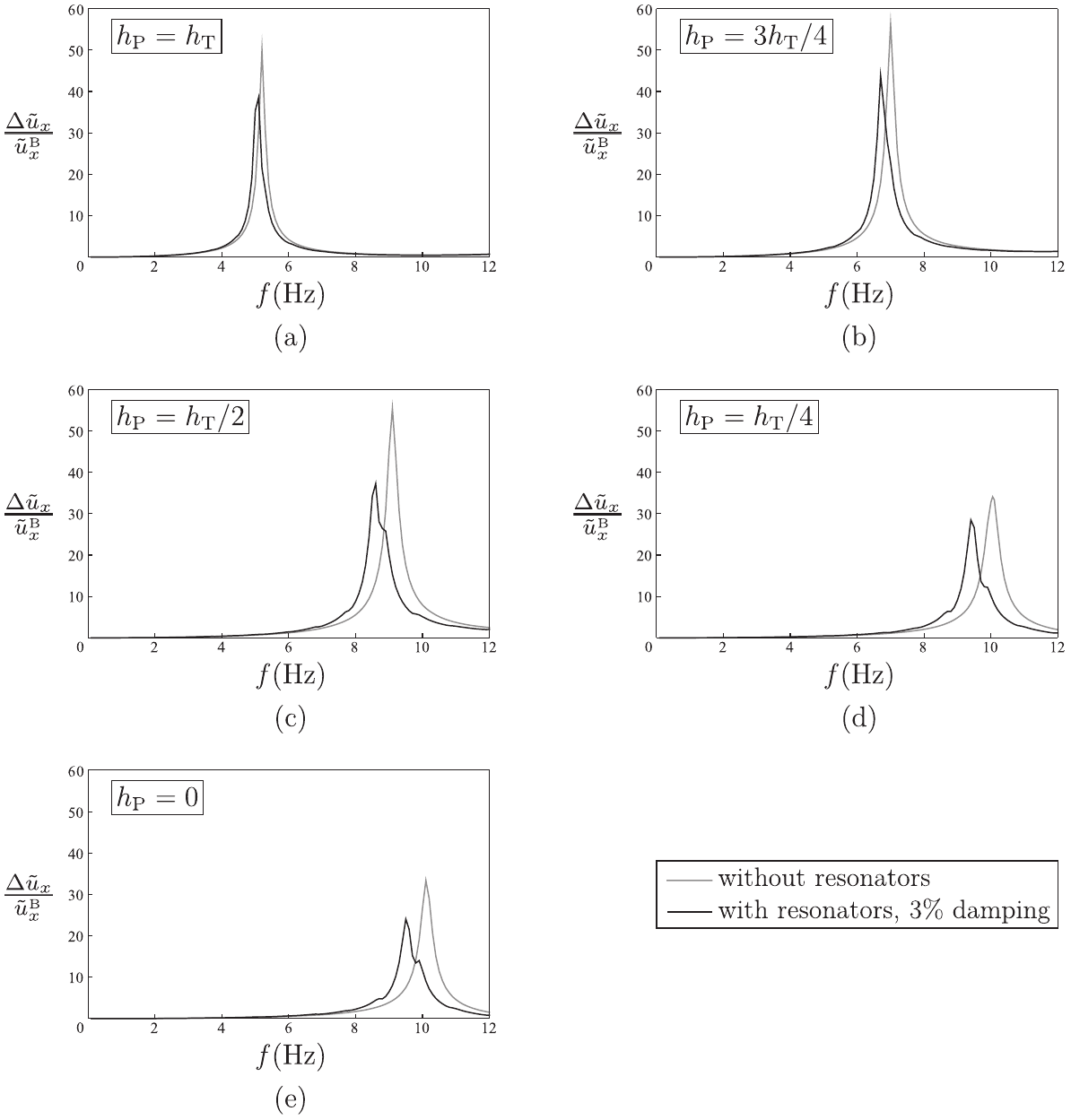}
\caption[Response of the fluid-filled tank with different resonators]{Same as in Fig.~\ref{FrequencyDomain1}, but with lighter resonators (\emph{design 2}).}
\label{FrequencyDomain2}
\end{figure}

Since the total mass of the resonators used in \emph{design 1} could be too large in terms of construction process and cost, we determine the response of the system when \emph{design 2} is delivered.
The numerical results for the second alternative design are presented in Fig.~\ref{FrequencyDomain2}.
A comparison between Figs.~\ref{FrequencyDomain1} and~\ref{FrequencyDomain2} shows that a reduction in the mass of the system of resonators leads to a lower mitigation of the structural response of the tank, as expected.
Nonetheless, we will show in Section~\ref{Section3.3} that \emph{design 2} is effective in mitigating the vibrations of the fluid-filled container under real earthquakes.

\subsection[Transient behaviour of the fluid-solid system under seismic loads]{Transient behaviour of the fluid-solid system under seismic loads}
\label{Section3.3}

\begin{figure}[tp]
\centering
\includegraphics[width=1.0\columnwidth,keepaspectratio]{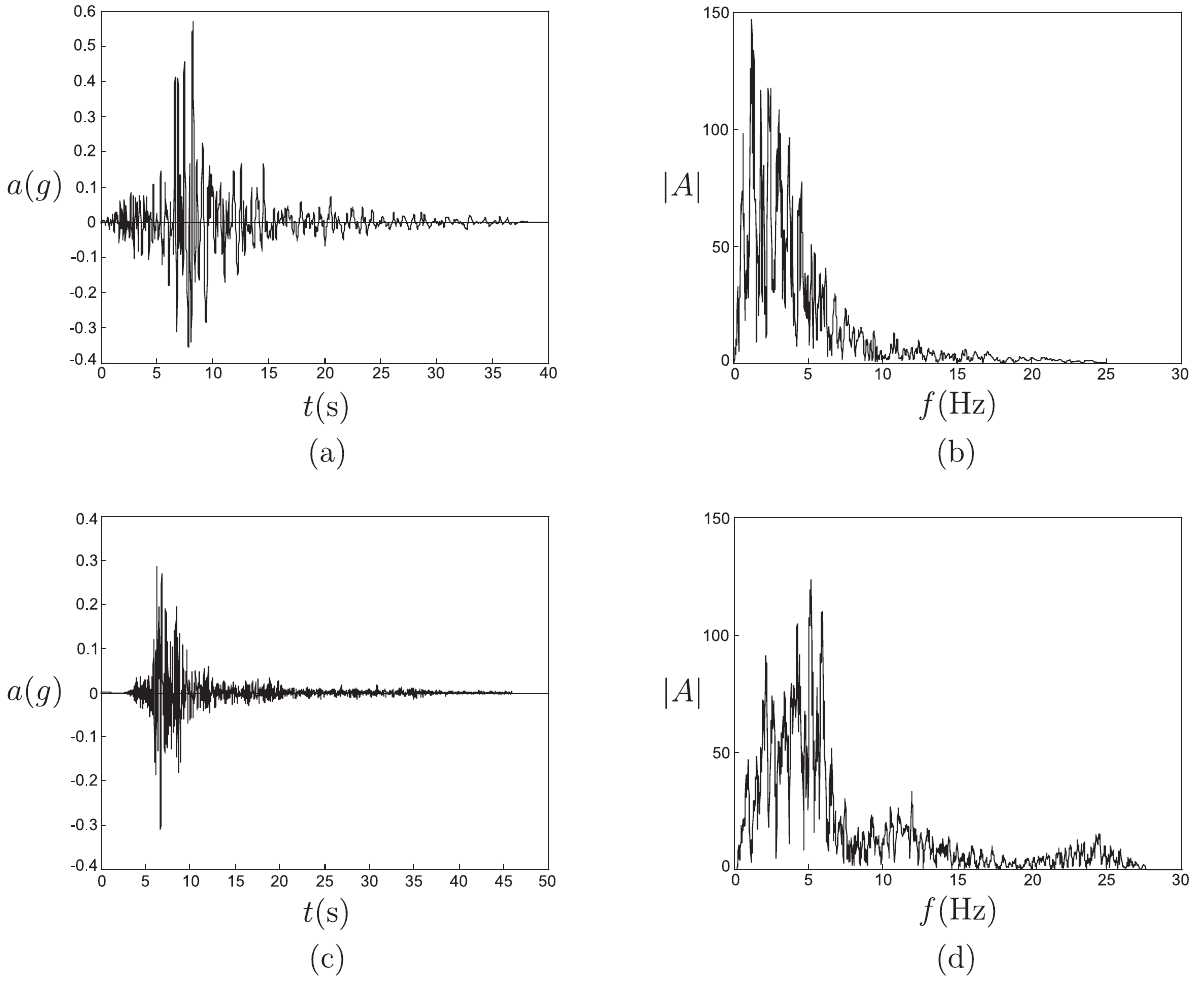}
\caption[Accelerograms used in computations]{Records (a,c) and relative Fourier amplitude spectra (b,d) of the accelerograms NORTHR\_ORR090.AT2 (Northridge, 1994) (a,b) and 001715YA (Ano Liosia, 1999) (c,d).}
\label{Earthquakes}
\end{figure}

In this Section, we compute the transient response of the fluid-filled tank when subjected to real seismic excitations.
The first accelerogram, labelled NORTHR\_ORR090.AT2 and plotted in Fig.~\ref{Earthquakes}a, represents the $x$ component of the Northridge earthquake of magnitude \num{6.7}, recorded in~1994 at an epicentral distance of~\SI{20}{\kilo\metre}.
It was selected to excite the low-frequency components of the tank response, close to the sloshing modes in the tank.
Its frequency content, obtained with \emph{Matlab}\textsuperscript{\textregistered}, is shown in Fig.~\ref{Earthquakes}b.

Conversely, the second accelerogram, indicated by~001715YA and illustrated in Fig.~\ref{Earthquakes}c, defines the $y$ component of the Ano Liosia earthquake of magnitude \num{6.0}, recorded in~1999 at an epicentral distance of~\SI{14}{\kilo\metre}.
It was selected to excite the high-frequency components of the tank response, close to the impulsive behaviour of the tank.
The frequency content is shown in Fig.~\ref{Earthquakes}d.

We assume that the earthquakes act in the $x$ direction.
For each seismic input, we determine the relative displacement $\Delta u_{x} = u_{x}^{\textup{T}} - u_{x}^{\textup{B}}$ between the top and the bottom of the tank.
We consider two different levels of petrol: $h_{\textup{P}} = h_{\textup{T}}$ (fully filled) and $h_{\textup{P}} = 7 h_{\textup{T}}/8$, for which the natural frequencies of the fluid-solid system are close to the peaks in the amplitude spectra of the two earthquakes.
In the simulations performed in this Section, we implement the lighter resonators defined in \emph{design 2} (see Section~\ref{Section3.1.2}).

\begin{figure}[tp]
\centering
\includegraphics[width=0.925\columnwidth,keepaspectratio]{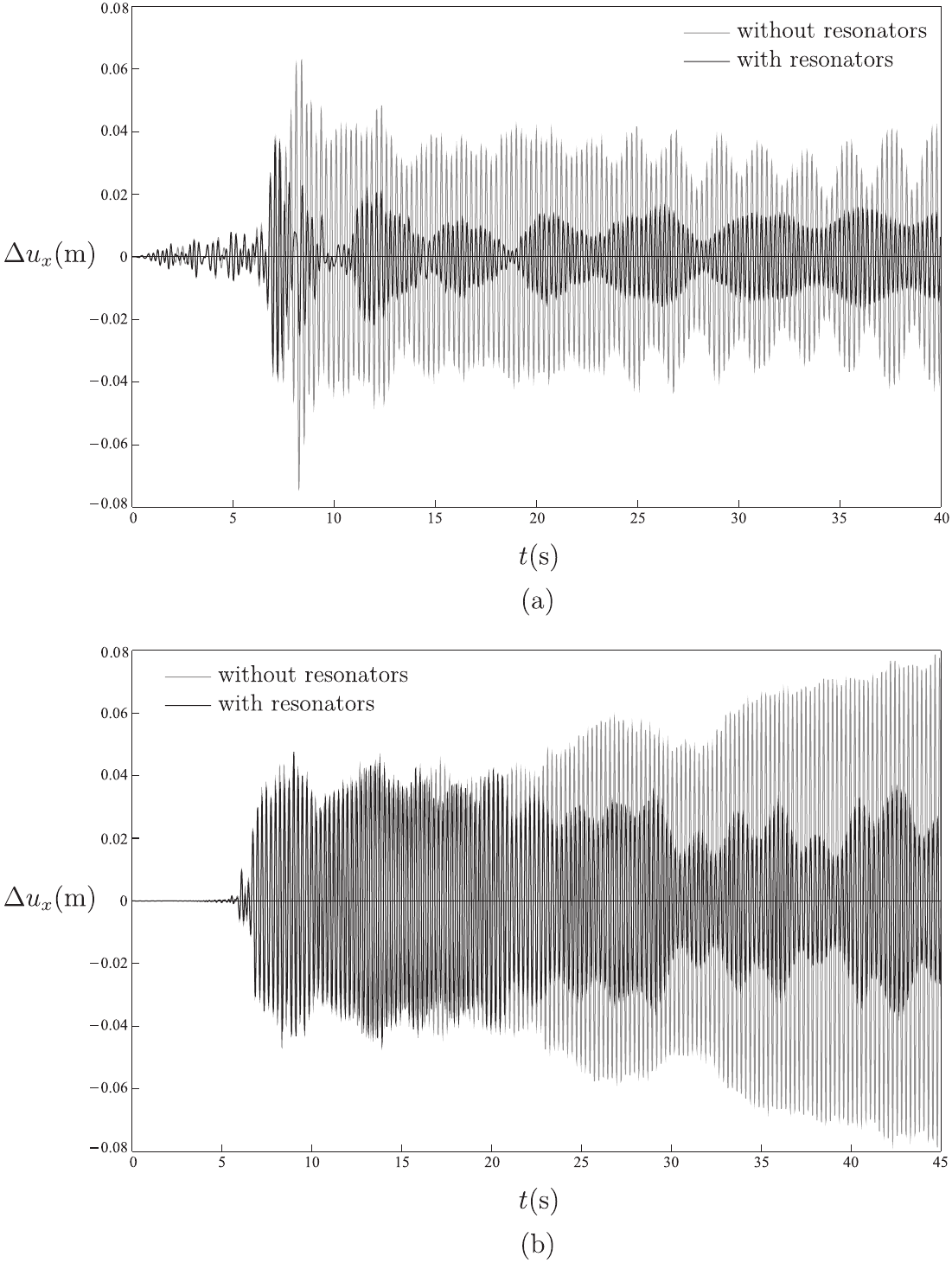}
\caption[Vibration of the storage tank with and without resonators under Northridge and Ano Liosia earthquakes]{Vibrations of the storage tank devoid of resonators (grey lines) and endowed with resonators (black lines), when acted upon by the Northridge earthquake (a) and by the Ano Liosia earthquake (b) (the properties of the resonators are chosen according to \emph{design 2}).}
\label{EarthquakesFluidLevel1}
\end{figure}

The grey lines in Figs.~\ref{EarthquakesFluidLevel1}a and~\ref{EarthquakesFluidLevel1}b represent the oscillations of the fully filled tank without resonators, under the two considered earthquakes.
If the resonators are installed in the system, the amplitudes of the oscillations of the container are decreased considerably, as indicated by the black lines.
More specifically, the reduction is of~\num{49.8}\% for the Northridge earthquake and of~\num{39.5}\% for the Ano Liosia earthquake.
Furthermore, resonant phenomena are prevented by employing the multi-scale resonators.

\begin{figure}[tp]
\centering
\includegraphics[width=0.925\columnwidth,keepaspectratio]{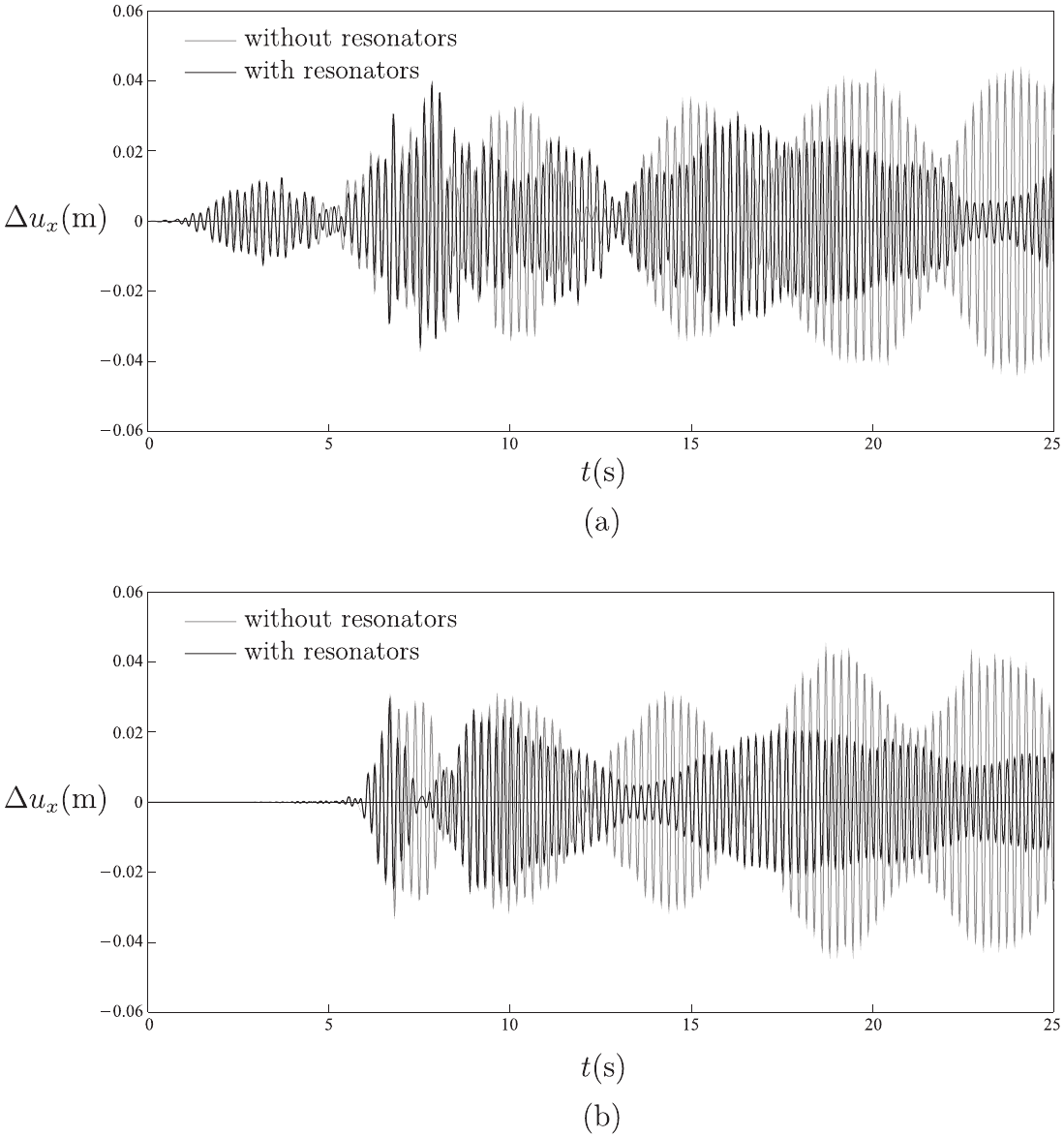}
\caption[Vibration of the storage tank with a different level of fluid]{Same as in Fig.~\ref{EarthquakesFluidLevel1}, but for a different level of fluid in the tank: $h_{\textup{P}} = 7 h_{\textup{T}}/8$.}
\label{EarthquakesFluidLevel2}
\end{figure}

The results for a different fluid level (i.e. $h_{\textup{P}} = 7 h_{\textup{T}}/8$) are illustrated in Fig.~\ref{EarthquakesFluidLevel2}.
In this case, the reduction is~\num{8.1}\% for the Northridge earthquake and~\num{34.1}\% for the Ano Liosia earthquake.
We have examined other fluid levels, the outcomes of which are not presented here for brevity, and we have found that in all cases the system of resonators attenuates the vibration amplitudes of the tank.
In some circumstances the resonators may not be very efficient, but the maximum attainable vibrations of the system -- occurring at its natural frequency -- are reduced if the resonators are attached to the structure, as shown in Section~\ref{Section3.2}.

\section[Waves in a large cluster of fluid-filled containers]{Waves in a large cluster of fluid-filled containers}
\label{Section4}

In an industrial facility some areas are usually covered by sets of fuel storage tanks, connected to each other by the foundation.
In regions of high seismic hazard, it is essential to study how these sets of tanks behave when they are subjected to an earthquake and how their vibrations can be reduced in order to avoid serious accidents due to structural failure.

For simplicity we look at two-dimensional systems, because numerical simulations with several three-dimensional cylindrical tanks would require large-scale computations.
Nonetheless, the ideas and the procedures presented in this Section can be easily extended to three-dimensional geometries.

\subsection[Dispersion analysis for a periodic system of containers]{Dispersion analysis for a periodic system of containers}
\label{Section4.1}

A set of many tanks can be modelled as a periodic system, consisting of an infinite array of repetitive cells.
The analysis of this system can be simplified by studying a single cell with Floquet-Bloch boundary conditions at the boundaries.
In this way, we can determine the dispersive properties of the system, in particular the frequency ranges where waves propagate without attenuation (\emph{pass-bands}) and where they decay exponentially in space (\emph{stop-bands}).

\subsubsection[Periodic cell and dispersion curves]{Periodic cell and dispersion curves}
\label{Section4.1.1}

\begin{figure}[tp]
\centering
\includegraphics[width=1.0\columnwidth,keepaspectratio]{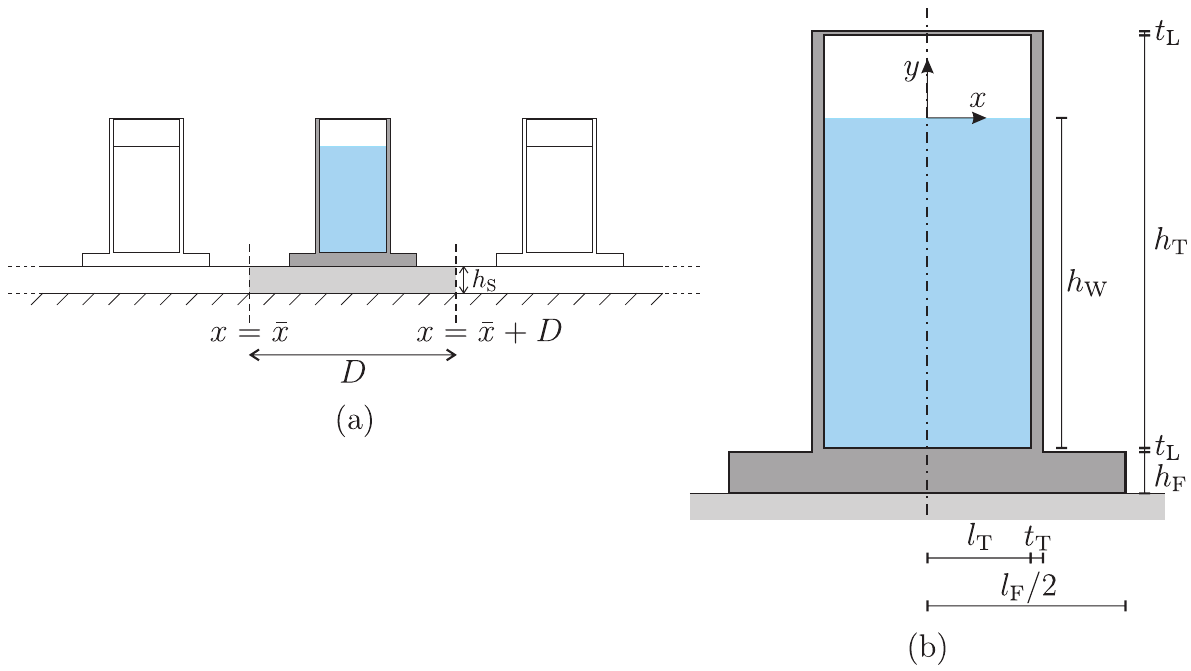}
\caption[Sketch of the periodic cell]{(a) Identification of the repetitive cell in the periodic system of containers; (b) geometrical details of each container.}
\label{PeriodicSystem}
\end{figure}

The repetitive cell of the periodic system of containers is indicated in Fig.~\ref{PeriodicSystem}a.
We assume that each container has a rectangular section and unit thickness in the $z$ direction, orthogonal to the $xy$ plane (see Fig.~\ref{PeriodicSystem}b).
The container is rigidly connected to a lid at the top and to a rectangular foundation at the bottom.
In the computations, the height of the fluid $h_{\textup{W}}$ will be varied.
Furthermore, we assume that the containers lie on a medium-dense sand soil, which stands over a rigid layer of rock; hence, we impose zero vertical displacements at the bottom of the sand layer.

At the ends of the periodic cell we apply Floquet-Bloch conditions:
\begin{equation}\label{FBconditions}
\tilde{\bm{u}}\left(\bar{x}+D,y\right) = \tilde{\bm{u}}\left(\bar{x},y\right) e^{\uimmi k D} \, ,
\end{equation}
where $\tilde{\bm{u}}$ is the displacement field, $k$ is the wavenumber and $D$ is the distance between two adjacent cells.

The dispersion curves of the periodic system, which show how the frequency $f$ depends on the wavenumber $k$, are determined numerically by means of a finite element analysis under plain strain conditions.
In particular, they are obtained by computing the eigenfrequencies of the periodic cell for several values of the wavenumber, ranging within the first Brillouin zone.

\subsubsection[Dispersion diagrams in absence of resonators - Numerical example]{Dispersion diagrams in absence of resonators - Numerical study}
\label{Section4.1.2}

We assume that the tank, the lid and the foundation are made of concrete, having Young's modulus $E_{\textup{T}} = \SI{30}{\giga\pascal}$, Poisson's ratio $\nu_{\textup{T}} = \num{0.2}$ and density $\rho_{\textup{T}} = \SI{2500}{\kilogram.\metre^{-3}}$.
The height and the half-width of the container are given by $h_{\textup{T}} = \SI{10}{\metre}$ and $l_{\textup{T}} = \SI{2.5}{\metre}$, respectively, while the thickness of the walls is $t_{\textup{T}} = \SI{0.3}{\metre}$.
The lid has thickness $t_{\textup{L}} = \SI{0.1}{\metre}$.
The foundation has a rectangular cross-section with height $h_{\textup{F}} = \SI{1.0}{\metre}$ and width $l_{\textup{F}} = \SI{9.6}{\metre}$.
The fluid is assumed to have the properties of water, with density $\rho_{\textup{W}} = \SI{1000}{\kilogram.\metre^{-3}}$, bulk modulus $K_{\textup{W}} = \SI{2.2}{\giga\pascal}$ and dynamic viscosity $\mu_{\textup{W}} = \SI{0.001}{\pascal.\second}$.
The sand soil has a thickness $h_{\textup{S}} = \SI{2}{\metre}$ and is characterised by a Young's modulus $E_{\textup{S}} = \SI{70}{\mega\pascal}$, a Poisson's ratio $\nu_{\textup{S}} = \num{0.25}$ and a density $\rho_{\textup{S}} = \SI{2100}{\kilogram.\metre^{-3}}$.
The length of the periodic cell is taken as $D = \SI{15.6}{\metre}$.

\begin{figure}[tp]
\centering
\includegraphics[width=1.0\columnwidth,keepaspectratio]{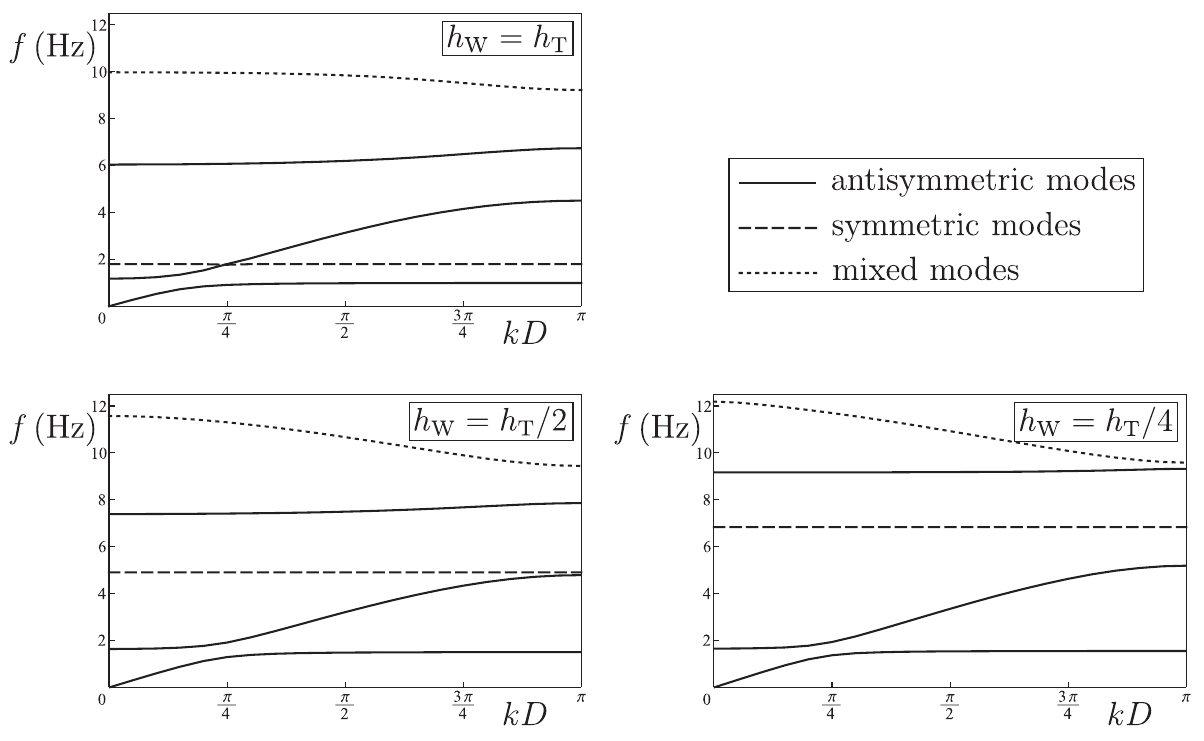}
\caption[Dispersion curves for the periodic system of containers without resonators]{Dispersion curves for the periodic system of containers without resonators depicted in Fig.~\ref{PeriodicSystem}, computed for three different levels of fluid.}
\label{DispersionCurves}
\end{figure}

We consider three different levels of fluid, i.e. $h_{\textup{W}} = h_{\textup{T}}$, $h_{\textup{W}} = h_{\textup{T}}/2$ and $h_{\textup{W}} = h_{\textup{T}}/4$, where $h_{\textup{W}}$ and $h_{\textup{T}}$ denote the heights of the fluid and of the container respectively.
The dispersion curves corresponding to these three cases are plotted in Fig.~\ref{DispersionCurves}, where the solid lines represent the antisymmetric modes, the dashed lines the symmetric modes and the dotted lines the mixed modes.
For each case, the dispersion curves associated with symmetric modes are almost flat, while those associated with mixed modes appear at higher frequencies.

The results in Fig.~\ref{DispersionCurves} show that, as the height of fluid is decreased, the dispersion curves move up towards higher frequencies.
This result can be explained from a physical point of view considering that the eigenfrequencies of a system generally become larger if its mass is diminished.
At the boundaries of the first Brillouin zone standing waves are observed, where the group velocity is zero.
We also note that, for each case examined, there are large finite frequency intervals between the dispersion curves, which represent the stop-bands of the system, where waves cannot propagate.

\subsubsection[Effects of the multi-scale resonators on the dispersion diagrams]{Effects of the multi-scale resonators on the dispersion diagrams}
\label{Section4.1.3}

As shown in Fig.~\ref{DispersionCurves}, the dispersion curves of the periodic system of containers can vary significantly with the level of fluid inside the containers.
Therefore, in order to suppress or reduce the structural vibrations, we would need an isolation device that is effective in a large frequency interval.
This purpose cannot be achieved with conventional Tuned Mass Dampers, which work well only in a narrow frequency range around the natural frequency of the damper.
On the other hand, the system of resonators described in Section~\ref{Section2.2} is capable of mitigating the vibrations of the containers for many frequencies within a large interval, defined by the dispersion relation\eqref{DispersionRelation}.

We design the resonators in order to cover the interval $[0,10]\si{\hertz}$, where the dispersion curves related to the antisymmetric modes of the system of containers are found for very different values of fluid heights (see Fig.~\ref{DispersionCurves}).
For this system, we attach to each container two chains of nine resonators, designed as steel spheres of diameter~\SI{0.35}{\metre} and connected by steel beams of length~\SI{1}{\metre} and circular cross-section of diameter~\SI{0.035}{\metre}.
The Young's modulus, Poisson's ratio and density of steel are~\SI{200}{\giga\pascal}, \num{0.3} and~\SI{7850}{\kilogram.\metre^{-3}}, respectively.
The system of resonators is linked to the top of the container by a rigid truss and tied to the foundation by a hinge.
We point out that the mass of each beam is much smaller than the mass of each resonator, which justifies the approximation, used in the derivation of Eq.\eqref{DispersionRelation}, that the beams can be analysed as non-inertial flexural elements.
Furthermore, we note that the total mass of the system of resonators is less than~\num{3}\% of the total mass of the fluid-filled container.

\begin{figure}[tp]
\centering
\includegraphics[width=1.0\columnwidth,keepaspectratio]{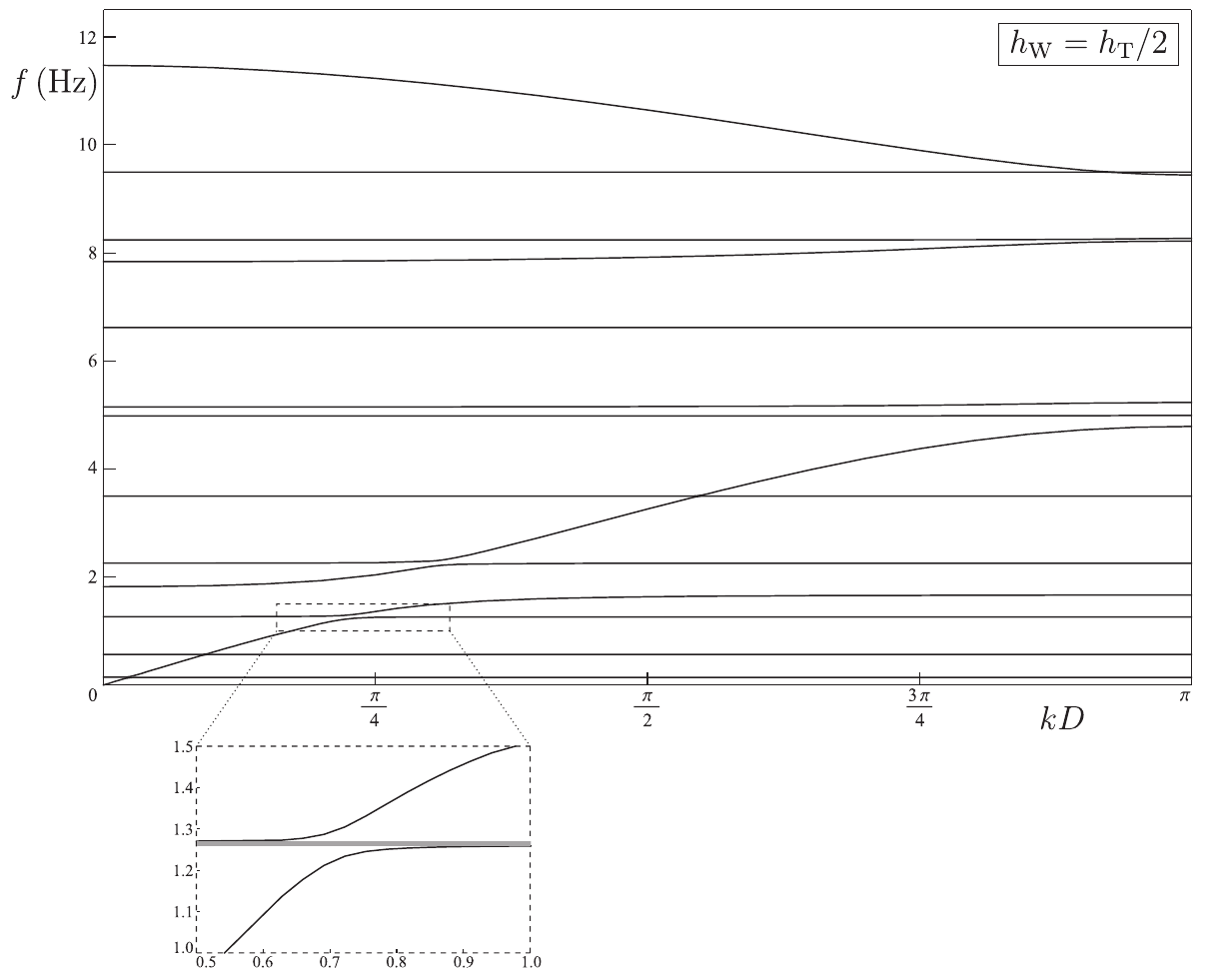}
\caption[Dispersion curves for the periodic system of containers with resonators]{Dispersion curves for the periodic system of containers when the resonators are installed in the system, for the case $h_{\textup{W}} = h_{\textup{T}}/2$.
The inset shows more clearly a stop-band created by the resonators, highlighted in grey.}
\label{DispersionCurvesResonators}
\end{figure}

If we connect the system of resonators to the container and we perform the dispersion analysis, we find that new dispersion curves appear in correspondence with the modes of the resonators.
These dispersion curves lie within the frequency interval $[0,10]\si{\hertz}$, in agreement with the design target.
In Fig.~\ref{DispersionCurvesResonators} we present the results for the case $h_{\textup{W}} = h_{\textup{T}}/2$, namely when the containers are half-filled with fluid.
Narrow stop-bands open up in proximity of the new dispersion curves, where waves decay exponentially; furthermore, standing waves are detected at the limits of the pass-bands.

\subsection[Transient response of a large cluster of containers]{Transient response of a large cluster of containers}
\label{Section4.2}

\begin{figure}[tp]
\centering
\includegraphics[width=1.0\columnwidth,keepaspectratio]{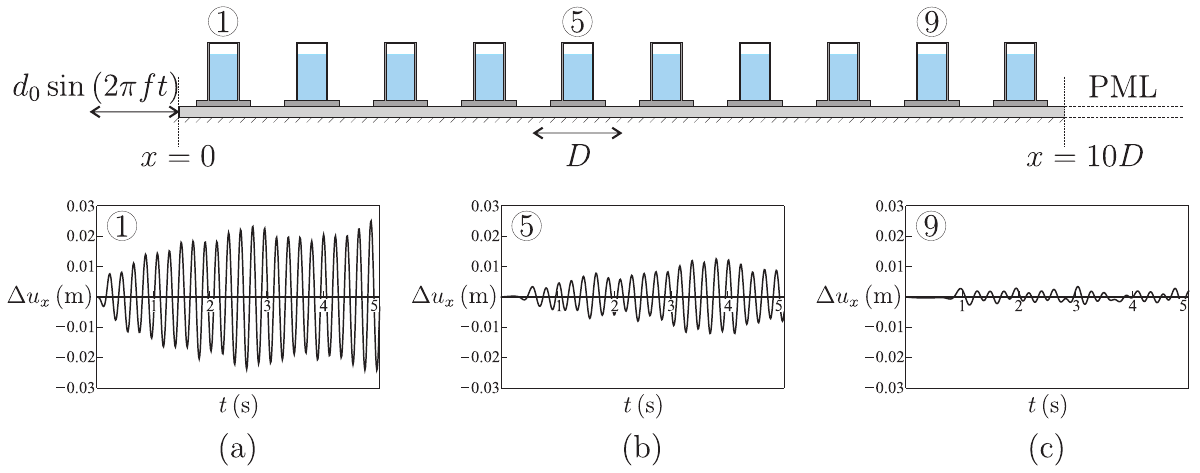}
\caption[Time-histories of the relative displacements between the top and the bottom of the first, fifth and ninth container]{Time-histories of the relative displacements between the top and the bottom of the first, fifth and ninth containers of a semi-infinite set of containers, shown on top, evaluated near the limit of a stop-band ($f = \SI{4.79}{\hertz}$).}
\label{Stop-Band}
\end{figure}

We consider a set of ten containers connected by the soil layer, sketched at the top of Fig.~\ref{Stop-Band}.
At the left boundary of the system ($x = 0$) we impose a harmonic horizontal displacement, given by $d_{0} \, \sin{(2 \pi f t)}$, with $d_{0} = \SI{0.002}{\metre}$ and for different values of the frequency $f$.
At the right boundary of the system ($x = 10 D$) we attach a soil layer with viscous damping, which absorbs waves.
In this way, reflection of waves at the right boundary of the system is avoided and the system behaves as semi-infinite.
The layer with damping plays the role of a \emph{Perfectly Matched Layer} (PML), in which the damping coefficient has been tuned in order to minimise reflections, as discussed in~\cite{Carta2014c}.

First, we consider an applied frequency equal to $f = \SI{4.79}{\hertz}$, which lies at the beginning of a stop-band (see Fig.~\ref{DispersionCurves}).
We determine the relative displacements $\Delta u_{x}$ between the top and the bottom of the first, fifth and ninth containers, which are reported at the bottom of Fig.~\ref{Stop-Band}.
It is apparent that the amplitude of $\Delta u_{x}$ decreases considerably moving from the first to the ninth container, as expected since waves decay exponentially in space when the frequency is inside a stop-band.
Accordingly, in this case resonators are not needed.

\begin{figure}[tp]
\centering
\includegraphics[width=1.0\columnwidth,keepaspectratio]{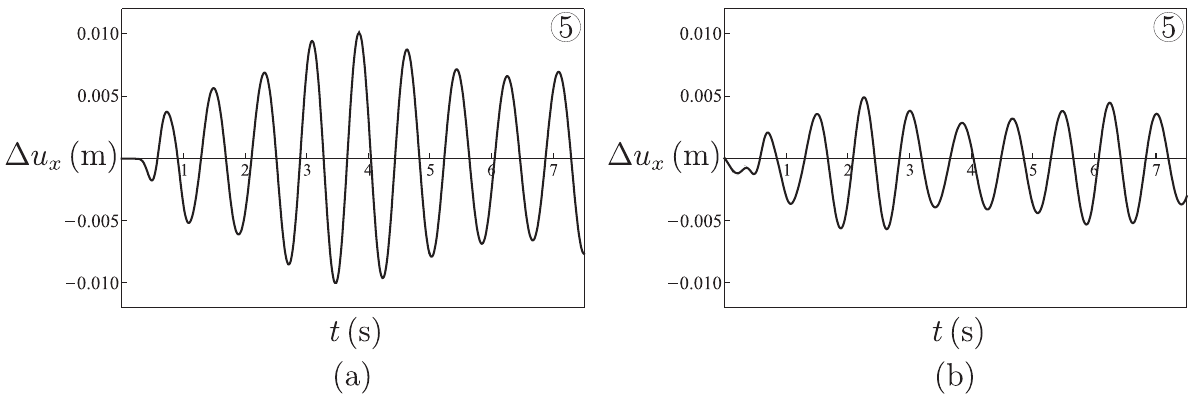}
\caption[Time-histories of the relative displacements between the top and the bottom of the fifth container]{Time-histories of the relative displacements between the top and the bottom of the fifth container in a semi-infinite system without resonators (a) and with resonators (b), calculated at $f = \SI{1.25}{\hertz}$.}
\label{TransientResponseSystem}
\end{figure}

Next, we impose a harmonic horizontal displacement at $x = 0$ with a frequency $f = \SI{1.25}{\hertz}$, falling within a pass-band of the system (see Fig.~\ref{DispersionCurves}).
In Fig.~\ref{TransientResponseSystem}a we plot the time-history of the relative displacement between the top and the bottom of the fifth container when the resonators are absent, while in Fig.~\ref{TransientResponseSystem}b we illustrate the response of the fifth container after the resonators have been installed in the system.
The comparison between the two cases shows that the resonators can reduce the structural vibrations of around~\num{50}\%.

\begin{figure}[tp]
\centering
\includegraphics[width=1.0\columnwidth,keepaspectratio]{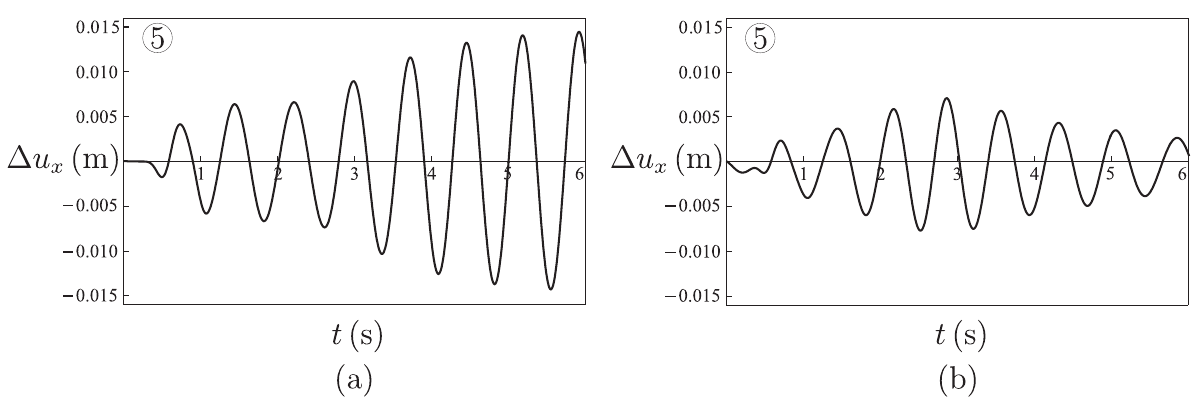}
\caption[Time-histories of the relative displacements between the top and the bottom of the fifth container at different frequency]{Same as in Fig.~\ref{TransientResponseSystem}, but computed at $f = \SI{1.35}{\hertz}$.}
\label{TransientResponseSystem2}
\end{figure}

The chosen value of the frequency is close to one of the flat dispersion curves produced by the resonators (see Fig.~\ref{DispersionCurvesResonators}).
Nonetheless, attenuation of waves has been observed also for other frequencies close to $f = \SI{1.25}{\hertz}$, namely at $f = \SI{1.35}{\hertz}$ (as shown in Fig.~\ref{TransientResponseSystem2}) and $f = \SI{1.15}{\hertz}$.

\section[Sloshing waves in the transient regime]{Sloshing waves in the transient regime}
\label{Section5}

Mathematical modelling of waves in fluids is a classical subject, which has generated a lot of interest among applied mathematicians, physicists and engineers.
In particular, the classical texts~\cite{Ursell1958,Ursell1994} and~\cite{Kuznetsov2002} present an excellent theoretical framework of the theory of water waves in the context of partial differential equations.
The dynamics of sloshing is well described in~\cite{Ibrahim2005}, which provides elegant estimates of eigenfrequencies of sloshing waves in solid containers.

In this Section, we investigate the influence of the multi-scale resonators, attached to the cylindrical tank, on  sloshing waves.

The behaviour of sloshing waves in systems endowed with isolation devices has been studied extensively in the literature.
In~\cite{ChaKel1988,ChaKel1990} it was observed that base-isolation provides a reduction in the dynamic response of a tank but, on the other hand, the amplitudes of sloshing waves are slightly amplified.
By means of finite element simulations, in~\cite{Malhotra1997} and~\cite{Gregoriou2005} the effects of base-isolators on the tank and on the sloshing were investigated, leading to results that are in agreement with the observations in~\cite{ChaKel1988,ChaKel1990}; namely the stress in the structure is reduced and buckling of the shell is avoided, but the the sloshing is slightly amplified when isolators are applied.
The same observations were obtained in~\cite{Cho-Kim-Lim-Cho:2004} with more complex analyses of base-isolated systems, taking into account liquid-structure-soil interaction.
In~\cite{ChrWhi2008} it was showed that the wave heights in the fluid are unaffected by the introduction of a seismic isolation system, but their analysis is limited to the case of a transient study with \lq standard' earthquake acceleration spectra and does not include the transient response of the system under a harmonic loading with a frequency equal to the first sloshing frequency.

Differently from base-isolators, the system of resonators we propose does not increase the sloshing wave amplitudes, even in proximity of the frequencies of sloshing modes, as demonstrated in the following.

\subsection[Sloshing waves in the two-dimensional container]{Sloshing waves in the two-dimensional container}
\label{Section5.1}

\begin{figure}[tp]
\centering
\includegraphics[width=0.9\columnwidth,keepaspectratio]{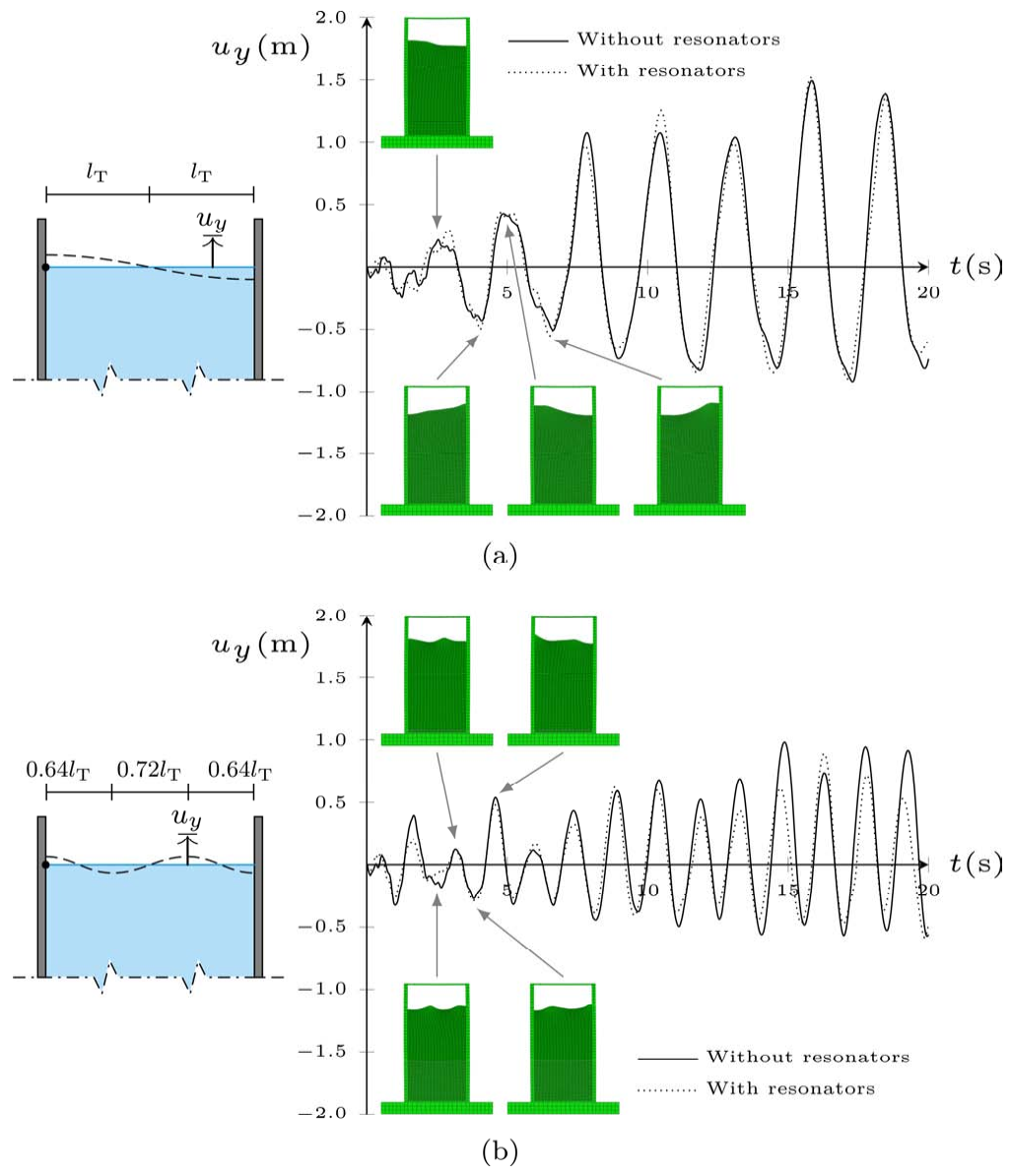}
\caption[Time-histories of the vertical displacement of the fluid]{Time-histories of the vertical displacement $u_{y}$ of the fluid point indicated by the black dot in the schematic representation on the left, computed at the frequencies~\SI{0.395}{\hertz} (a) and~\SI{0.684}{\hertz} (b).
The solid and dotted lines refer to the cases without and with resonators, respectively.}
\label{ResultsSloshingWaves2D}
\end{figure}

First, we carry out numerical simulations in the transient regime for the two-dimensional container examined in Section~\ref{Section4}.
We impose a sinusoidal displacement at the bottom of the foundation, having a frequency equal to one of the frequencies of sloshing waves.
Then, we determine the vertical displacement $u_{y}$ of a point of the fluid close to the boundary with the solid, indicated by a black dot in the insets of Fig.~\ref{ResultsSloshingWaves2D}.
In the computations, we assume that the fluid level is $h_{\textup{W}} = \SI{8}{\metre}$.

The time-histories of $u_{y}$ at~\SI{0.395}{\hertz} and~\SI{0.684}{\hertz}, which are the frequencies of the first and second antisymmetric modes of sloshing waves (see~\ref{AppendixB.2}), are plotted by solid lines in Fig.~\ref{ResultsSloshingWaves2D}a and~\ref{ResultsSloshingWaves2D}b, respectively.
The amplitudes of the applied displacement are~\SI{0.05}{\metre} and~\SI{0.04}{\metre}, respectively.
For these simulations, the finite element software \emph{Abaqus}\textsuperscript{\textregistered} has been used.
The deformations of the fluid at different instants of time are shown in the insets of the diagrams, where the shapes of the first and second modes are clearly identified.
For both frequencies, the level of the fluid increases unbounded with time due to resonance.

The dotted lines in Figs.~\ref{ResultsSloshingWaves2D}a and~\ref{ResultsSloshingWaves2D}b represent the time-histories of $u_{y}$ when the resonators are installed in the system.
It is apparent that the resonators do not significantly modify the amplitudes of sloshing waves, therefore they result to be inefficacious to reduce the sloshing in the fluid.
The reason is that at these frequencies the displacements of the container walls are very small, so that the container behaves as a rigid body and the resonators cannot divert the energy from the fluid.
However, we note that the resonators do not amplify the sloshing in the fluid, as occurs in a tank with base-isolation.

\subsection[Sloshing waves in the three-dimensional tank]{Sloshing waves in the three-dimensional tank}
\label{Section5.2}

\begin{figure}[tp]
\centering
\includegraphics[width=1.0\columnwidth,keepaspectratio]{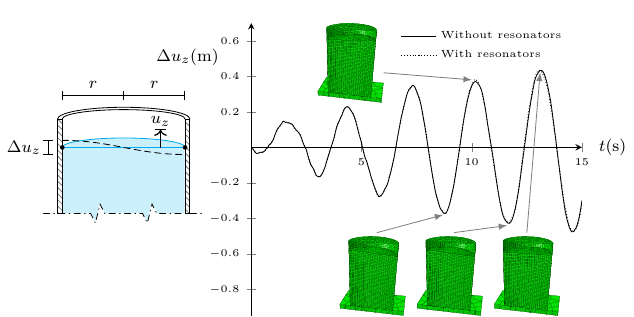}
\caption[Time-histories of the vertical displacement of the fluid]{Time-histories of the difference $\Delta u_{z}$ between the vertical displacements of the fluid points indicated by the black dots in the schematic representation on the left, computed at the frequency~\SI{0.338}{\hertz}.
The solid and dotted lines refer to the cases without and with resonators, respectively.}
\label{ResultsSloshingWaves3D}
\end{figure}

We perform transient regime simulations also for the three-dimensional container examined in Sections~\ref{Section2} and~\ref{Section3}.
We take the fluid level as $h_{\textup{P}} = \SI{12}{\metre}$.
We impose a sinusoidal displacement at the bottom of the foundation, having a frequency equal to~\SI{0.338}{\hertz}, which corresponds to the first antisymmetric mode of sloshing waves (see \ref{AppendixB.1}), and an amplitude equal to~\SI{0.02}{\metre}.
Using the finite element software \emph{Abaqus}\textsuperscript{\textregistered}, we compute the difference between the vertical displacements $u_{z}$ of the two points of the fluid free surface closer to the boundary with the solid, indicated by the black dots in the inset of Fig.~\ref{ResultsSloshingWaves3D}.
The time-history of the difference $\Delta u_{z}$ is plotted by a solid line in Fig.~\ref{ResultsSloshingWaves3D}.

The deformations of the fluid at different instants of time are shown in the insets of the diagram, where the shapes of the first mode are clearly identified.
Moreover, we can note that the level of the fluid increases unbounded with time due to resonance.

The dotted line in Fig.~\ref{ResultsSloshingWaves3D} represents the time-history of $\Delta u_{z}$ when the resonators are installed in the system.
Similarly to the two-dimensional case, we note that at this frequency the container behaves as a rigid body and the resonators do not significantly modify the amplitudes of sloshing waves.
The simplest solution to reduce the sloshing in the fluid would be to introduce baffles (Belakroum et al.~(2010), Wang et al.~(2012)), which do not affect the properties of the resonators.

Finally, we observe that if the structure were more flexible, the frequencies of sloshing waves and the eigenfrequencies of the filled-fluid container would be closer.
In this case, sloshing waves could increase the vibrations of the structure and viceversa.
In such a system, the resonators would be very useful, as they would reduce the oscillations of the container and, as a consequence, the amplitudes of sloshing waves would not be increased by the structural vibrations.

\section[Conclusions]{Conclusions}
\label{Section6}

The paper has presented an innovative design for isolation of vibrations of multi-scale structures consisting of fluid-filled containers.
The idea of employing high-contrast multi-scale resonators has proved to be elegant and efficient to reduce vibrations of elastic containers filled with fluid within a predefined interval of frequencies.

The analytical estimates for the choice of parameters of the resonators have been based on the Floquet-Bloch approach, which provides a constructive guidance for the design of the multi-scale resonators.
The numerical simulations have been carried out in the full fluid-solid interaction mode, with the fluid velocity and pressure satisfying the Navier-Stokes equations.
Frequency response analyses have been performed in the framework of the linear water wave theory, and their results have been confirmed by transient simulations both in two and three dimensions.

In the practical implementation, viscous energy dissipating dampers are desirable and can be attached to the multi-scale resonators system.
The current design clearly identifies the ``energy sink'', and channels the energy of vibrations away from the main body of the fuel tank.
The combined system, which includes viscous dampers, complements the design to the level of a robust structure.

The mass of the resonators and the stiffness of the connecting flexural links can be changed without affecting the dispersion properties of the system of resonators, as described in the analytical model presented in the main text of the paper.
Hence, depending on the power of the external waves, the resonators can be chosen to be heavier or lighter, in order to absorb the required amount of energy from the vibrating fuel container.

The model has high potential impact on industrial applications, concerning in particular the protection of existing fuel storage tanks in petrochemical plants subjected to seismic waves.
However, the applicability of the model is not limited to fluid-filled tanks in a seismic region, but can be extended to the protection of a wide range of multi-scale structures subjected to different dynamic loads.

\section*{Acknowledgments}
\noindent
G.C. and O.S.B. acknowledge the support from the Research Fund for Coal and Steel of the European Commission, INDUSE-2-SAFETY project, grant number RFSR-CT-2014-00025.
A.B.M. would like to thank the EPSRC (UK) for its support through Programme grant no. EP/L024926/1.
L.P.A. would like to thank the University of Liverpool for financial support and provision of excellent research facilities.
The final part of the work was completed while A.B.M. was visiting the University of Trento, with the support from ERC Advanced Grant ‘Instabilities and nonlocal multi-scale modelling of materials’ FP7-PEOPLE-IDEAS-ERC-2013-AdG; this support is gratefully acknowledged.


\appendix

\section[Eigenfrequencies and eigenmodes of the fluid-filled tanks]{Eigenfrequencies and eigenmodes of the fluid-filled tanks}
\label{AppendixA}
\setcounter{figure}{0}
\renewcommand{\thefigure}{A.\arabic{figure}}

The eigenfrequencies and eigenmodes of the fluid-solid systems are computed by means of finite element models developed in \emph{Comsol Multiphysics}\textsuperscript{\textregistered}, assuming in each case that the bottom of the foundation is fixed ($\bm{u}_{\textup{F}}=\bm{0}$).
We focus the attention on the antisymmetric modes of the systems, since these are the only modes excited by a horizontal acceleration of the ground~\cite{GraRod1952}.

\subsection[Three-dimensional tank]{Three-dimensional tank}
\label{AppendixA.1}

\begin{figure}[tp]
\centering
\includegraphics[width=0.75\columnwidth,keepaspectratio]{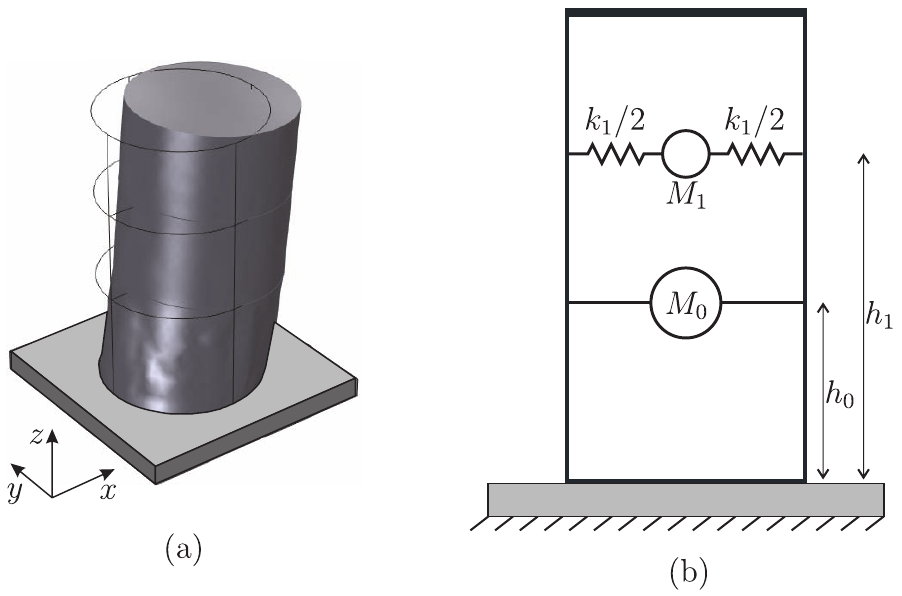}
\caption[Three-dimensional fluid-filled tank antisymmetric eigenmodes and spring-mass approximation]{(a) First antisymmetric eigenmode of the three-dimensional fluid-filled tank; (b) spring-mass approximation of the cylindrical storage tank.}
\label{Eigenmodes3D}
\end{figure}

First, we perform the eigenfrequency analysis by linearising the equations~\eqref{EquationsFluidTH} of the fluid and removing the viscous terms (\emph{linear water wave theory}).
This approach is the most precise, as both the pressure and the velocity of the fluid are used as variables of the problem.
As an example, we consider a fluid level $h_{\textup{P}} = \SI{12}{\metre}$, while the other constitutive and geometric properties of the system are the same as in Section~\ref{Section3.1}.
The first antisymmetric eigenmode is shown in Fig.~\ref{Eigenmodes3D}a, and it corresponds to an eigenfrequency $f_{1} = \SI{6.11}{\hertz}$.

If the fluid is instead modelled as an acoustic medium, the first eigenfrequency associated with an antisymmetric mode is found to be $f_{1}^{\textup{ac}} = \SI{6.15}{\hertz}$, which is very close to the value determined with the linear water wave theory.
The reason is that the frequencies of sloshing waves are considerably smaller than~\SI{6.15}{\hertz} (see~\ref{AppendixB.1}), therefore the contribution of the fluid velocity can be ignored.
On the other hand, at low frequencies the model based on the linearised Navier-Stokes equations should be employed, since in this case the velocity of sloshing waves is significant.

An alternative approach consists in representing the fluid as a a system of masses connected to the main structure by springs.
This method, which is referred to as \emph{spring-mass approximation}, is commonly used in practical applications, because it is simple and requires a lower computational cost.
For the three-dimensional tank, we refer to the papers~\cite{Housner1957,Housner1963} and~\cite{Veletsos:1984}.
The fluid is represented by a mass $M_{0}$, rigidly connected to the container, and by a mass $M_{1}$, linked to the walls by an elastic spring of equivalent stiffness $k_{1}$ (see Fig.~\ref{Eigenmodes3D}b).
The positions of the masses with respect to the bottom of the tank are indicated by $h_{0}$ and $h_{1}$, respectively.
The expressions of these quantities are reported below:
\begin{subequations}\label{EquationsSpringMass3D}
\begin{align}
M_0 &= \frac{\tanh{\left( \sqrt{3} \frac{r}{h_{\textup{P}}} \right)}}{\sqrt{3} \frac{r}{h_{\textup{P}}}} M \, , \label{EquationsSpringMass3DM0} \\
h_0 &= \frac{9}{20}h_{\textup{P}} & & \text{for} \quad \frac{h_{\textup{P}}}{r}>\frac{8}{3} \, , \label{EquationsSpringMass3Dh0} \\
M_1 &= \left(\frac{11}{24}\right)^2 \sqrt{\frac{27}{8}}\frac{r}{h_{\textup{P}}} \tanh{\left( \sqrt{\frac{27}{8}}\frac{h_{\textup{P}}}{r} \right)} M \, , \label{EquationsSpringMass3DM1} \\
k_1 &= \frac{g}{r} \sqrt{\frac{27}{8}} \tanh{\left( \sqrt{\frac{27}{8}}\frac{h_{\textup{P}}}{r} \right)} M_1 \, , \label{EquationsSpringMass3Dk1} \\
h_1 &= \left[ 1 - \frac{\cosh{\left( \sqrt{\frac{27}{8}}\frac{h_{\textup{P}}}{r} \right)}-\frac{135}{88}}{\sqrt{\frac{27}{8}}\frac{h_{\textup{P}}}{r} \sinh{\left( \sqrt{\frac{27}{8}}\frac{h_{\textup{P}}}{r} \right)}} \right] h_{\textup{P}} \, , \label{EquationsSpringMass3Dh1}
\end{align}
\end{subequations}
where $M = \pi r^{2} h_{\textup{P}} \rho_{\textup{P}}$ is the total mass of the fluid.
For the same choice of the parameters considered above, the first antisymmetric mode is found at $f_{1}^{\textup{sm}} = \SI{6.44}{\hertz}$.
This value of the eigenfrequency is larger than the values determined with the more accurate methods based on the linear water wave theory and the acoustic approximation.
Therefore, the spring-mass approximation should be employed only to have a rough estimate of the first eigenfrequency of the three-dimensional fluid-filled tank.

\subsection[Two-dimensional container]{Two-dimensional container}
\label{AppendixA.2}

\begin{figure}[tp]
\centering
\includegraphics[width=1.0\columnwidth,keepaspectratio]{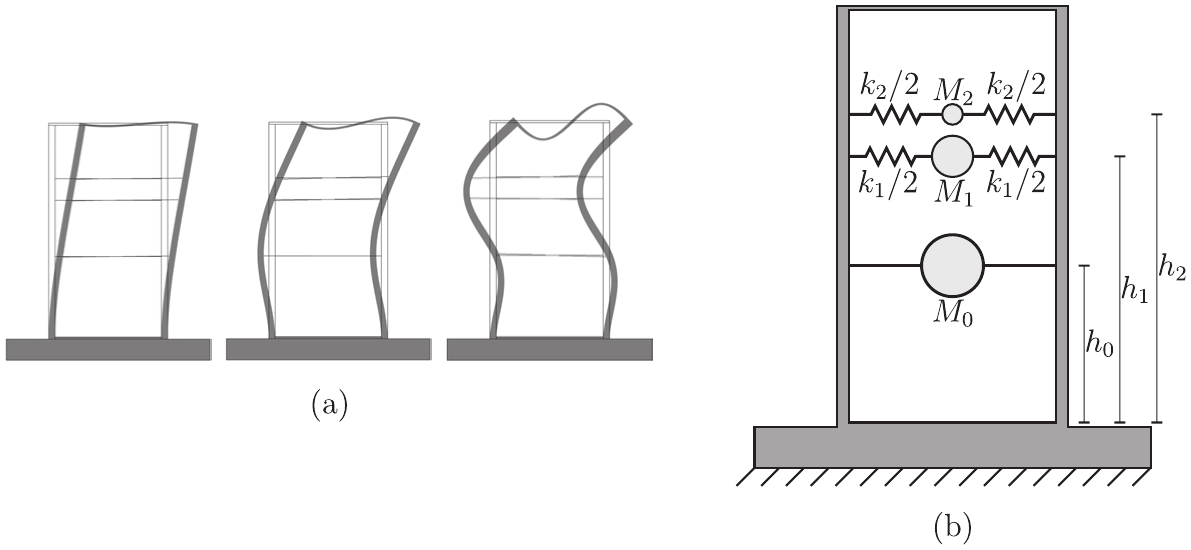}
\caption[Two-dimensional fluid-filled tank antisymmetric eigenmodes and spring-mass approximation]{(a) First three antisymmetric eigenmodes of the two-dimensional fluid-filled container; (b) spring-mass approximation of the container.}
\label{Eigenmodes2D}
\end{figure}

In this case we take the fluid level equal to $h_{\textup{W}} = \SI{8}{\metre}$, while the other properties are identical to those in Section~\ref{Section4.1.2}.
The first three antisymmetric eigenmodes obtained from the linear water wave theory are shown in Fig.~\ref{Eigenmodes2D}a.
The associated eigenfrequencies are given by $f_{1} = \SI{1.28}{\hertz}$, $f_{2} = \SI{6.17}{\hertz}$ and $f_{3} = \SI{17.25}{\hertz}$, respectively.

If the fluid is substituted by an acoustic medium, the first three eigenfrequencies associated with antisymmetric eigenmodes are $f_{1}^{\textup{ac}} = \SI{1.29}{\hertz}$, $f_{2}^{\textup{ac}} = \SI{6.18}{\hertz}$ and $f_{3}^{\textup{ac}} = \SI{17.30}{\hertz}$.
Also in the two-dimensional case, the acoustic approximation provides a good estimate of the natural frequencies of the fluid-solid system.
This is due to the fact that the first resonant modes of sloshing waves are found at much lower values of the frequency (refer to~\ref{AppendixB.2} for details), consequently the contribution of fluid flow at higher frequencies is negligible.

The spring-mass approximation has been applied to two-dimensional containers in the works~\cite{GraRod1952}, \cite{Housner1957}, \cite{Li2012} and~\cite{Bursi2016a}, just to cite a few examples.
For the rectangular containers examined in Section~\ref{Section4} we consider the model in~\cite{LiWang2012}, which is an improvement of the formulation provided in~\cite{GraRod1952}.
The fluid is represented by a mass $M_{0}$, rigidly connected to the container, and by a set of spring-mass oscillators $(M_{n},k_{n})$ ($n = 1,2,\dots$) that simulate the sloshing modes of the fluid.
We consider the case of two spring-mass oscillators, sketched in Fig.~\ref{Eigenmodes2D}b.
The positions of the masses $M_{0}$, $M_{1}$ and $M_{2}$ with respect to the bottom of the tank are indicated by $h_{0}$, $h_{1}$ and $h_{2}$, respectively.
The formulae for all the quantities indicated in Fig.~\ref{Eigenmodes2D}b are reported below:
\begin{subequations}\label{EquationsSpringMass2D}
\begin{gather}
M_n = \frac{2(h_{\textup{W}}/l_{\textup{T}})^2 \tanh{(\beta_n h_{\textup{W}})}}{(\beta_n h_{\textup{W}})^3} M \quad \left( n=1,2 \right) \, , \label{EquationsSpringMass2DMn} \\
h_n = \left[ 1 + \frac{2-\cosh{(\beta_n h_{\textup{W}})}}{\beta_n h_{\textup{W}}\sinh{(\beta_n h_{\textup{W}})}} \right] h_{\textup{W}} \quad \hfill \left( n=1,2 \right) \, , \label{EquationsSpringMass2Dhn} \\
M_0 = \left[ 1 - \sum_{n=1}^{2} \frac{2(h_{\textup{W}}/l_{\textup{T}})^2 \tanh{(\beta_n h_{\textup{W}})}}{(\beta_n h_{\textup{W}})^3} \right] M \, , \label{EquationsSpringMass2DM0} \\
h_0 = \frac{\frac{1}{2}+\frac{1}{3}\left(\frac{l_{\textup{T}}}{h_{\textup{W}}}\right)^2-2\left(\frac{h_{\textup{W}}}{l_{\textup{T}}}\right)^2 \sum_{n=1}^{2}\frac{2+\beta_n h_{\textup{W}} \sinh{(\beta_n h_{\textup{W}})}-\cosh{(\beta_n h_{\textup{W}})}}{(\beta_n h_{\textup{W}})^4 \cosh{(\beta_n h_{\textup{W}})}}}{1 - \sum_{n=1}^{2} \frac{2(h_{\textup{W}}/l_{\textup{T}})^2 \tanh{(\beta_n h_{\textup{W}})}}{(\beta_n h_{\textup{W}})^3}} h_{\textup{W}} \, , \label{EquationsSpringMass2Dh0} \\
k_n = M_n \omega_{(n)}^2 = \frac{2 M g}{h_{\textup{W}}} \left(\frac{h_{\textup{W}}}{l_{\textup{T}}}\right)^2 \left[\frac{\tanh{(\beta_n h_{\textup{W}})}}{\beta_n h_{\textup{W}}}\right]^2 \quad \left( n=1,2 \right) \, . \label{EquationsSpringMass2Dkn}
\end{gather}
\end{subequations}
Here $\beta_{n} = (2n-1)\pi/(2 l_{\textup{T}})$, $\omega_{(n)}$ is the $n$-th frequency of the sloshing waves, and $M = 2 l_{\textup{T}} h_{\textup{W}} \rho_{\textup{W}}$ is the total mass per unit thickness of the fluid.
Taking the same values of the system properties as above, the first three eigenfrequencies corresponding to antisymmetric eigenmodes that are obtained with the approximate spring-mass model are $f_{1}^{\textup{sm}} = \SI{1.39}{\hertz}$, $f_{2}^{\textup{sm}} = \SI{4.85}{\hertz}$ and $f_{3}^{\textup{sm}} = \SI{20.45}{\hertz}$, respectively.
A comparison with the results of the linear water wave theory and of the acoustic formulation shows that the spring-mass model does not provide a good estimate of the natural frequencies of the fluid-solid system, in particular for the second and third modes.
We have also checked that the accuracy of this approximation does not improve significantly by increasing the number of spring-mass oscillators.

\section[Analytical study of the frequencies of sloshing waves]{Analytical study of the frequencies of sloshing waves}
\label{AppendixB}
\setcounter{figure}{0}
\renewcommand{\thefigure}{B.\arabic{figure}}

The frequencies of sloshing waves in the fluid are calculated analytically in the framework of linear water wave theory, assuming that the walls of the container are rigid.
The continuity equation and the linearised Navier-Stokes equations are given by
\begin{subequations}\label{EquationsFluidBis}
\begin{gather}
\nabla \cdot \bm{v}_{\textup{f}} = 0 \, , \label{ContinuityBis} \\
\frac{\partial \bm{v}_{\textup{f}}}{\partial t} + \frac{\nabla p}{\rho_{\textup{f}}} = -g \, \bm{k} \, . \label{LinearisedNavierStokes}
\end{gather}
\end{subequations}
We recall that $\rho_{\textup{f}}$, $\bm{v}_{\textup{f}}$, $p$, $g$ and $\bm{k}$ are the density of the fluid, the velocity field, the pressure, the acceleration of gravity and the unit vector in the direction normal to the free surface, respectively.
The boundary conditions in a rigid container are expressed by
\begin{subequations}\label{BoundaryConditionsFluidRigid}
\begin{align}
\bm{v}_{\textup{f}} \cdot \bm{n} &= 0 & & \text{on the rigid walls} , \label{BoundaryConditionsFluidRigid1} \\
p &= p_{0} & & \text{on the free surface} , \label{BoundaryConditionsFluidRigid2}
\end{align}
\end{subequations}
where $p_{0}$ is the atmospheric pressure and $\bm{n}$ is the unit vector normal to the rigid walls.

Considering an irrotational flow, it is possible to define a velocity potential $\varphi$ such that $\bm{v}_{\textup{f}} = \nabla \varphi$.
Assuming small perturbations of the free surface, the equations of the system and the boundary conditions become (see also~\cite{Ibrahim2005}):
\begin{subequations}\label{EquationsFluidTris}
\begin{align}
\nabla^{2} \varphi &= 0 & & \text{inside the fluid} , \label{EquationsFluidTris1} \\
\nabla \varphi \cdot \bm{n} &= 0 & & \text{on the rigid walls} , \label{EquationsFluidTris2} \\
\nabla \varphi \cdot \bm{n} &= \frac{\omega^{2} \varphi}{g} & & \text{on the free surface} , \label{EquationsFluidTris3}
\end{align}
\end{subequations}
where $\omega$ is the radian frequency.
The solution of the linear water waves problem, defined by Eqs.~\eqref{EquationsFluidTris}, provides the radian frequencies of sloshing waves.
We are interested, in particular, in the frequencies of sloshing waves relative to antisymmetric modes, which are the only modes that are excited when the fluid-solid system is subjected to ground shaking.

\subsection[Three-dimensional tank]{Three-dimensional tank}
\label{AppendixB.1}

The radian frequencies of sloshing waves in the three-dimensional cylindrical tank examined in Sections~\ref{Section2} and~\ref{Section3} are given in~\cite{Ibrahim2005}:
\begin{equation}\label{FrequenciesSloshingWaves3D}
\omega_{(mn)}^{\textup{3D}} = \sqrt{\frac{g \xi_{mn}}{r} \tanh{\left( \frac{\xi_{mn} h_{\textup{P}}}{r} \right)} } \quad \text{with} \quad \xi_{mn} = \lambda_{mn} r \quad \left( m,n = 1,2,\dots \right) \, ,
\end{equation}
where $r$ is the radius of the cylinder, $h_{\textup{P}}$ is the height of the fluid and $\lambda_{mn}$ are the roots of the equation
\begin{equation}\label{BesselEquation}
\left. \frac{\partial J_{m} \left( \lambda_{mn}\rho \right)}{\partial \rho} \right|_{\rho=r} = 0 \, ,
\end{equation}
being $J_{m}$ the Bessel function of the first kind of order $m$.
With the values assigned to the parameters in Section~\ref{Section3.1} and for $h_{\textup{P}} = \SI{12}{\metre}$, the first three frequencies of sloshing waves related to antisymmetric modes are found to be $f_{(11)}^{\textup{3D}} = \SI{0.338}{\hertz}$, $f_{(31)}^{\textup{3D}} = \SI{0.511}{\hertz}$ and $f_{(12)}^{\textup{3D}} = \SI{0.576}{\hertz}$, where $f_{(mn)}^{\textup{3D}} = \omega_{(mn)}^{\textup{3D}}/(2 \pi)$.
The same values have been obtained from a finite element model developed in \emph{Comsol Multiphysics}\textsuperscript{\textregistered}.

\subsection[Two-dimensional container]{Two-dimensional container}
\label{AppendixB.2}

For what concerns the two-dimensional rectangular container studied in Section~\ref{Section4}, the radian frequencies have the following expressions~\cite{Ibrahim2005,LiWang2012}:
\begin{equation}\label{FrequenciesSloshingWaves2D}
\omega_{(n)}^{\textup{2D}} = \sqrt{g \, \beta_n \tanh{(\beta_n h_{\textup{W}})}} \quad \text{with} \quad \beta_n = \frac{n \pi}{2 l_{\textup{T}}} \quad \left( n = 1,2,\dots \right) \, ,
\end{equation}
where $l_{\textup{T}}$ and $h_{\textup{W}}$ are the half-width of the container and the height of the fluid, respectively.
Using the values of the parameters chosen in Section~\ref{Section4.1.2} and for $h_{\textup{W}} = \SI{8}{\metre}$, the first three frequencies of sloshing waves associated with antisymmetric modes result to be $f_{(1)}^{\textup{2D}} = \SI{0.3951}{\hertz}$, $f_{(3)}^{\textup{2D}} = \SI{0.6844}{\hertz}$ and $f_{(5)}^{\textup{2D}} = \SI{0.8835}{\hertz}$, where $f_{(n)}^{\textup{2D}} = \omega_{(n)}^{\textup{2D}} / (2 \pi)$.
Also in this case, the finite element model built in \emph{Comsol Multiphysics}\textsuperscript{\textregistered} provides a very accurate estimate of the analytical results.

If we consider an elastic container, the frequencies of sloshing waves decrease very slightly (as also found experimentally by Jiang et al.~(2014)).
The reason is that, both in the two-dimensional and in the three-dimensional cases, the deformations of the container are very small at those frequencies.
Accordingly, we have not considered the frequencies of sloshing waves as eigenfrequencies of the fluid-solid system.

\end{document}